\documentclass[prl,amsmath,amssymb,aps,superscriptaddress,floatfix,twocolumn,notitlepage]{revtex4-1}
\usepackage{graphicx}
\usepackage{color}
\usepackage{dcolumn}
\usepackage{bm}
\usepackage[utf8]{inputenc}
\usepackage{amssymb}
\usepackage{color,amsmath}
\usepackage{mathtools}
\usepackage{soul,xcolor} 
\setstcolor{red} 
\usepackage{epstopdf}
\DeclareUnicodeCharacter{03BD}{{$\nu$}}
\usepackage{hyperref}
\hypersetup{
 pdfnewwindow=true, colorlinks=true,
 linkcolor=blue, anchorcolor=blue,
 citecolor=blue, filecolor=blue,
 menucolor=blue, urlcolor=blue}

\begin{document}
\title{Isospin Pomeranchuk effect and the entropy of collective excitations in twisted bilayer graphene}
\author{Yu Saito}
\thanks{These authors contributed equally}
\affiliation{Department of Physics, University of California at Santa Barbara, Santa Barbara, CA 93106, USA}
\affiliation{California NanoSystems Institute, University of California at Santa Barbara, CA 93106 USA}
\author{Fangyuan Yang}
\thanks{These authors contributed equally}
\affiliation{Department of Physics, University of California at Santa Barbara, Santa Barbara, CA 93106, USA}
\author{Jingyuan Ge}
\affiliation{Department of Physics, University of California at Santa Barbara, Santa Barbara, CA 93106, USA}
\author{Xiaoxue Liu}
\affiliation{Department of Physics, Brown University, Providence, RI 02912, USA}
\author{Takashi Taniguchi}
\affiliation{International Center for Materials Nanoarchitectonics,
National Institute for Materials Science,  1-1 Namiki, Tsukuba 305-0044, Japan}
\author{Kenji Watanabe}
\affiliation{Research Center for Functional Materials,
National Institute for Materials Science, 1-1 Namiki, Tsukuba 305-0044, Japan}
\author{J.I.A. Li}
\affiliation{Department of Physics, Brown University, Providence, RI 02912, USA}
\author{Erez Berg}
\affiliation{Department of Condensed Matter Physics, Weizmann Institute of Science, Rehovot 76100, Israel}
\author{Andrea F. Young}
 \email{andrea@physics.ucsb.edu}
\affiliation{Department of Physics, University of California at Santa Barbara, Santa Barbara, CA 93106, USA}
\date{\today}
\begin{abstract}\end{abstract}
\maketitle

\textbf{In condensed matter systems, higher temperature typically disfavors ordered phases leading to an upper critical temperature for magnetism, superconductivity, and other phenomena.  
A notable exception is the Pomeranchuk effect in $^3$He, in which the liquid ground state freezes upon increasing the temperature\cite{pomeranchuk_theory_1950} due to the large entropy of the paramagnetic solid phase.  
Here we show that a similar mechanism describes the finite temperature dynamics of spin- and valley- isospins in magic-angle twisted bilayer graphene\cite{andrei_graphene_2020}. 
Most strikingly a resistivity peak appears at high temperatures near superlattice filling factor $\nu=-1$, despite no signs of a commensurate correlated phase appearing in the low temperature limit.  Tilted field magnetotransport and thermodynamic measurements of the in-plane magnetic moment show that the resistivity peak is connected to a finite-field magnetic phase transition\cite{lu_superconductors_2019} at which the system develops finite isospin polarization. These data are suggestive of a Pomeranchuk-type mechanism, in which the entropy of disordered isospin moments in the ferromagnetic phase stabilizes it relative to an isospin unpolarized Fermi liquid phase at elevated temperatures. 
Measurements of the entropy, $S/k_B$ indeed find it to be of order unity per unit cell area, with a measurable fraction that is suppressed by an in-plane magnetic field consistent with a contribution from disordered spins. In contrast to $^3$He, however, no discontinuities are observed in the thermodynamic quantities across this transition.
Our findings imply a small isospin stiffness\cite{wu_collective_2020,khalaf_charged_2020}, with implications for the nature of finite temperature transport\cite{polshyn_large_2019,cao_strange_2020,he_tunable_2020} as well as mechanisms underlying isospin ordering and superconductivity\cite{cao_unconventional_2018,yankowitz_tuning_2019} in twisted bilayer graphene and related systems.}

The best studied example of moir\'e flat band systems\cite{andrei_graphene_2020}, magic-angle twisted bilayer graphene\cite{bistritzer_moire_2011-1} is known to host a wide array of low-temperature phases\cite{cao_correlated_2018,cao_unconventional_2018,yankowitz_tuning_2019,sharpe_emergent_2019,lu_superconductors_2019,saito_independent_2020,stepanov_interplay_2019,arora_superconductivity_2020}. 
Some among these, notably correlated insulators\cite{cao_correlated_2018,yankowitz_tuning_2019,lu_superconductors_2019} and orbital Chern ferromagnets\cite{sharpe_emergent_2019,lu_superconductors_2019,serlin_intrinsic_2020,tschirhart_imaging_2020}, are known unambiguously to arise from the effects of the Coulomb interaction. However, the origin of other phenomena is less clear, particularly  superconductivity\cite{cao_unconventional_2018,yankowitz_tuning_2019,lu_superconductors_2019,saito_independent_2020,stepanov_interplay_2019,arora_superconductivity_2020} and the large scattering observed in the metallic finite temperature state\cite{polshyn_large_2019,cao_strange_2020}. The prevalence of Coulomb-driven phases is suggestive of a unified origin for all of the phenomenology.  In this picture both superconductivity and the finite temperature resistivity would arise from the interaction of charge carriers with collective modes of the spin and valley isospin.  
However, theoretical efforts based on conventional phononic mechanisms do appear to capture the basic experimental phenomenology of both the superconducting\cite{dodaro_phases_2018,wu_theory_2018,lian_twisted_2019,martin_moire_2020} and high-temperature metallic states\cite{wu_phonon-induced_2019}.  
We report on the discovery of an entropically driven phase transition between an isospin unpolarized Fermi liquid at low temperature and a state characterized by large, strongly fluctuating local magnetic moments at high temperature. Whereas no such phenomenology is expected from electron-phonon interaction alone, our results imply the existence of low-energy collective modes of electronic origin that couple strongly to the charge carriers. Such modes likely play a crucial role in determining the low-temperature phase diagram as well as the finite-temperature transport properties. 

\begin{figure*}[ht!]
\includegraphics[width= 135mm]{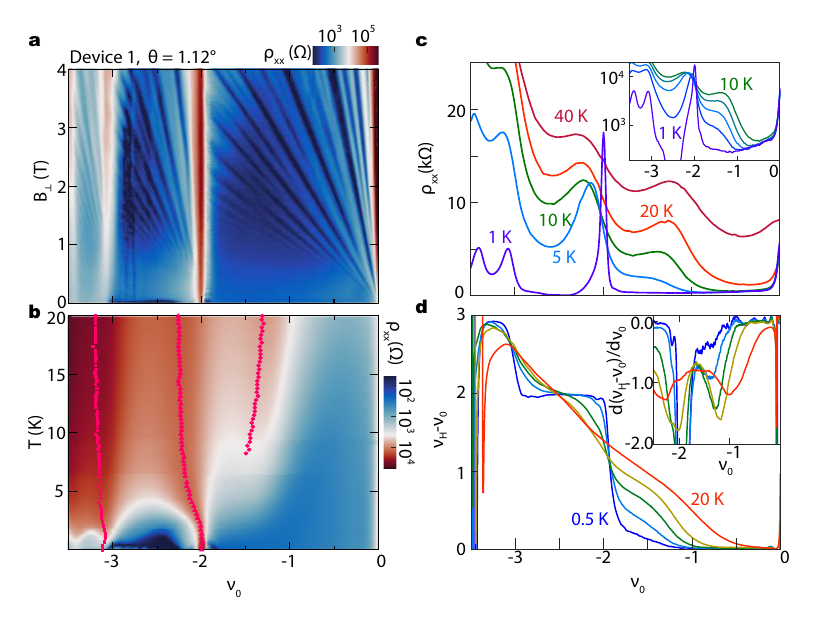}
 \caption{\textbf{Contrasting transport at low- and intermediate temperatures in twisted bilayer graphene near $\nu_0$ = -1.}
 \textbf{a}, Longitudinal resistivity $\rho_\mathrm{xx}$ in Device 1 as a function of the nominal superlattice filling factor $\nu_0$ and out-of-plane magnetic field $B_\perp$, acquired at $T$ = 400 mK. No correlated state is observed at $\nu_0=-1$
 Additional analysis of low-$B$ data, as well as data from Device 2, are shown in Fig. \ref{fig:low field fan} and \ref{fig:xx40_landaufan}.
 \textbf{b},   $\rho_\mathrm{xx}$ a function of temperature $T$ and $\nu_0$ at $B_\mathrm{tot} = 0$ T in Device 1. Pink circles correspond to the position of the local maxima in  $\rho_\mathrm{xx}$.  A resistivity peak emerges near $\nu_0=-1$ above T=5 K. 
\textbf{c}, $\rho_\mathrm{xx}$ traces at $T$ = 1, 5, 10, 20 and 40 K and $B_\mathrm{tot} = 0$ T. 
Inset: the same data plotted on a logarithmic scale.
 \textbf{d}, Hall density $\nu_\mathrm{H}-\nu_0$ as a function of $\nu_0$ at $T$ = 0.5, 2.5, 4.5, 8.0 and 20 K and $B_\mathrm{\perp}$ = 0.5 T. Inset: d$(\nu_\mathrm{H}-\nu_0)/d\nu_0$ as a function of $\nu_0$.
}
\label{fig:1}
\end{figure*}

Figure 1 shows transport measurements performed on a high quality twisted bilayer graphene device fabricated with inter-layer twist angle $\theta\approx 1.12^\circ$\cite{saito_hofstadter_2020} (see Fig. \ref{devices}).  
Figure \ref{fig:1}a shows the magnetoresistance measured at sub-kelvin temperatures, plotted as a function of the nominal electron filling $\nu_0$ of the superlattice unit cell and an applied out of plane magnetic field $B_\perp$. We determine $\nu_0$ from the geometric capacitance, measured from the Hall density near charge neutrality, and the positions of the most prominent resistivity features at $B_\perp = 0$ T which we associate with filling factors $\nu =-2,+2$, and $+3$.  
For $B_\perp$ of 1-2 T, distinct sets of quantum oscillations are observed at low magnetic fields that intersect the $B_\perp$=0 axis at $\nu_0=-3$,$-2$, and $0$ and show one-fold, two-fold, and four-fold degeneracy, respectively (see also Fig. \ref{fig:low field fan}). In graphene systems, spin and valley degeneracy typically give rise to quantum oscillations with degeneracy of four. The lower degeneracy of the quantum oscillations originating from $\nu_0=-2$ and $-3$ are consistent with ferromagnetism in which the ground states near those filling are polarized into one isospin flavor for $-4<\nu_0<-3$ or two flavors for $-3<\nu_0<-2$\cite{zondiner_cascade_2020}.  
No comparable `fan' is observed at low $B_\perp$ with intercept at $\nu_0=-1$, and the quantum oscillations originating from the charge neutrality point maintain apparent fourfold degeneracy for $-2<\nu_0<0$ suggesting unbroken flavor symmetry in that regime.  

\begin{figure*}[ht!]
\includegraphics[width=183mm]{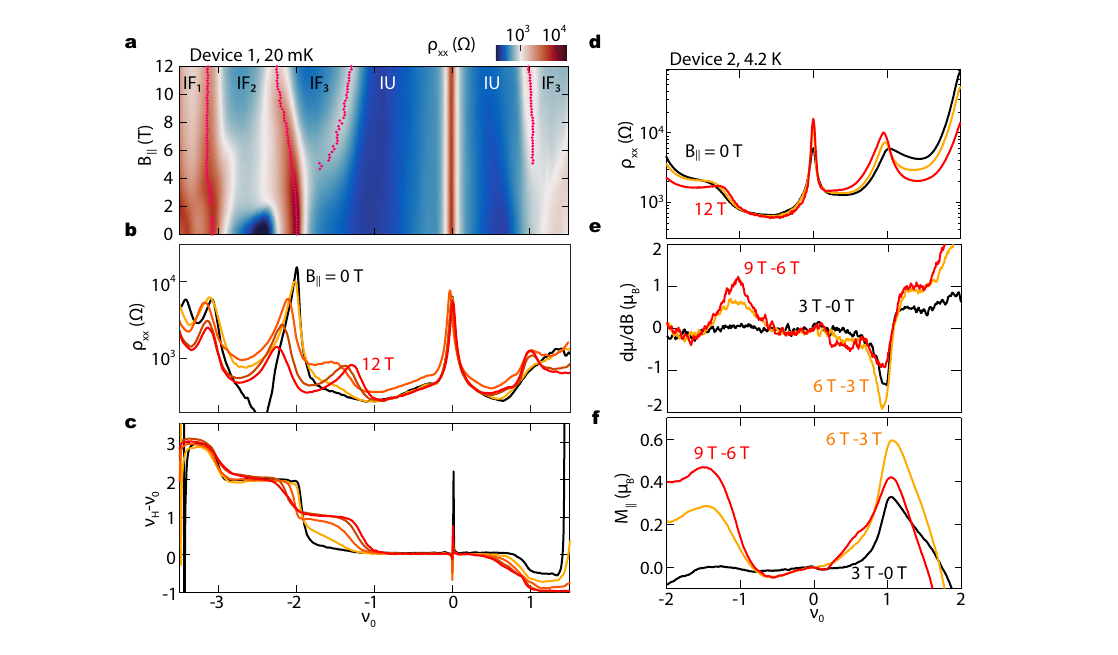}
 \caption{\textbf{In-plane magnetic field stabilized isospin ferromagnetism.}
\textbf{a}, $\rho_\mathrm{xx}$ as a function of $\nu_0$ and in-plane magnetic field $B_\parallel$ acquired at nominal temperature of $T$ = 20 mK in Device 1. Pink circles denote $\rho_\mathrm{xx}$ peak positions showing phase boundaries between symmetry breaking isospin ferromagnets (IF$_1$, IF$_2$, and IF$_3$) and an isospin unpolarized state (IU).
\textbf{b}, $\rho_\mathrm{xx}$ as a function of $\nu_0$ at $B_{\parallel}$ = 0, 3, 6, 9, 12 T.
\textbf{c} Subtracted Hall density $\nu_\mathrm{H}-\nu_0$ expressed in electrons per superlattice unit cell, and measured with $B_\mathrm{tot}$ = 0.5, 3, 6, 9, 12 T and fixed $B_\mathrm{\perp}$ = 0.5 T.
\textbf{d}, $\rho_\mathrm{xx}$ as a function of $\nu_0$ at $B_\parallel$ = 0, 6, 12 T and $T$ = 4.2 K in Device 2. 
\textbf{e}, $d\mu/dB$ as a function of $\nu_0$. Red, orange and black curves are calculated from finite differences between curves in panel d. 
\textbf{f}, Magnetization per superlattice unit cell as a function of $\nu_0$, obtained by integrating data in panel e with respect to $\nu_0$.
}
\label{fig:2}
\end{figure*}

The primary unexplained experimental phenomenology is illustrated in Figs. \ref{fig:1}b-c, which show transport data from the same device at higher temperatures.  Near $\nu_0=-3$ and $\nu_0=-2$, resistivity peaks associated with low-temperature correlated phases weaken and depin from commensurate $\nu$ as the temperature is raised.  In addition, they are joined by a third resistivity peak near---but not at---$\nu_0=-1$.  By 40 K, these peaks are indistinguishable, pointing to a universal behavior at this temperature.  We observe very similar behavior in a second device with $\theta=1.06^\circ$, (Device 2,  see Figs. \ref{fig:XX40_Rxx_map}, \ref{fig:Rxx_comp}, \ref{fig:vH_comp}, and \ref{fig:xx40_landaufan}).  Moreover, this behavior appears in the the published experimental literature on twisted bilayer graphene, where its origin has remained unexplained\cite{polshyn_large_2019,arora_superconductivity_2020}. Measurements of the Hall density show related behavior. Fig. \ref{fig:1}d  shows the Hall density $\nu_H$ plotted in units of electrons per unit cell and with the density arising from geometric capacitance ($\nu_0$) subtracted.  $\nu_H-\nu_0$ develops a pronounced kink between $\nu_0 =-1$ and $-2$ at high temperature similar to that seen near $\nu_0=-3$ and $-2$ (see also Fig. \ref{fig:vH_comp}). This is seen clearly in plots of $\frac{d}{d\nu_0}(\nu_H-\nu_0)$ (Fig. \ref{fig:1}d, inset), where the kinks appear as extrema.  Similar kinks have been associated with symmetry breaking at low temperatures\cite{xie_weak-field_2020}, specifically a reduction in the isospin symmetry of the Fermi surface.  The behavior of the Hall density and longitudinal resistivity suggest the existence of an isospin polarized phase at high temperatures that is absent in the zero temperature limit.   

The connection between high temperature resistivity peaks and isospin symmetry breaking is illustrated by the transport behavior at sub-kelvin temperatures in magnetic field $B_\parallel$ applied in the plane of the sample, which shows remarkably similar behavior to that at elevated temperature. 
As shown in Figs. \ref{fig:2}a-b, for $B_\parallel\gtrsim 3$ T, an additional resistance peak develops for $-2<\nu_0\lesssim-1$, while the resistance peak initially at $\nu_0=-2$ depins from this filling and decreases in magnitude as it moves to larger absolute $\nu_0$. 
The Hall density similarly shows the development of a new step near $\nu_0=-1$ that is absent at $B = 0$ T (Fig. \ref{fig:vH}). 
The behavior of the resistivity peak is roughly independent of the direction of the magnetic field orientation, showing nearly identical trajectories for in-plane or partially out-of-plane magnetic fields (Fig. \ref{all_landaufans_rxx}). 
Tilted field data  (Fig. \ref{tiltedLL_analysis} and \ref{fig:supple_LL_magni}) reveal that the resistivity peak separates discrete domains of quantum oscillations: for $\nu_0 >\nu_\mathrm{pk}$ (where $\nu_\mathrm{pk}$ denotes the filling at the peak in $-2<\nu_0\lesssim-1$) the oscillation minima remain qualitatively unchanged, extrapolating to $\nu_0=0$, while for $\nu_0 <\nu_\mathrm{pk}$ new quantum oscillations emerge that extrapolate to $\nu_0 = -1$ at $B = 0$ T.  We interpret the resistivity peak as a $B_\parallel$-driven transition from an isospin unpolarized (IU) paramagnetic state at low $B$ to a spin- and valley-polarized isospin ferromagnetic (IF$_3$) state at high $B$ in which electrons are polarized into three of four isospin flavors. This hypothesis is also consistent with the observation of a strong Chern insulator state with $C=\pm3$  in this regime in out-of-plane field\cite{saito_hofstadter_2020,wu_chern_2020,nuckolls_strongly_2020}. The additional resistivity peaks associated with $\nu_0 =-2$ and $\nu _0 =-3$ at low $T$ similarly denote boundaries between ferromagnetic phases (IF$_2$ and IF$_1$ phases) with fewer occupied isospin flavors.

Near $\nu=\pm1$, most proposed ordered ground states are expected to have finite spin polarization, making in-plane magnetization per unit cell,  $M_\parallel$, a good proxy for isospin polarization. To determine $M_\parallel$ we use Device 2\cite{liu_tuning_2020} whose geometry\cite{lee_chemical_2014,yang_experimental_2020} enables direct measurement of the chemical potential, $\mu$, of the twisted bilayer graphene layer (see Methods and Fig. \ref{devices}b). $M_\parallel$ is then extracted via the Maxwell relation 
$\left(\partial M_\parallel/\partial \nu\right)_{B_\parallel}=-\left(\partial \mu/\partial B_\parallel\right)_\nu$.
Transport is qualitatively similar between Devices 1 and 2; in particular, both devices show a $T$- or $B_\parallel$-induced resistance peak near $\nu_0=-1$ that is absent at low temperatures and field (Figs. \ref{fig:2}a, \ref{fig:XX40_Rxx_map} and \ref{fig:Rxx_comp}). In addition, the slightly smaller twist angle in Device 2 appears to favor a correlated state at $\nu_0 =+1$ that is absent at $B=0$ in Device 1 (Fig. \ref{fig:Rxx_comp}). 

$\partial \mu/\partial B_\parallel$ is shown in Fig. \ref{fig:2}e, determined using measurements of $\mu$ acquired at 3 T intervals.  The integrated $M_\parallel=-\int_0^{\nu_0} \frac{\partial \mu}{\partial B_\parallel} d\nu$ is shown in Fig. \ref{fig:2}f (see also Fig. \ref{chemical_potential}).  
Finite $M_\parallel$ is observed at $\nu_0 \approx 1$ even at the lowest magnetic fields in Device 2, consistent with the resistivity peak seen at the same filling being associated with an isospin ferromagnet with finite spin polarization. 
This buildup of magnetization near $\nu_0=1$ is consistent with prior observations~\cite{zondiner_cascade_2020,park_flavour_2020}. In contrast, near $\nu_0=-1$ no magnetization is observed in the measurement between $B_\parallel$=0 and 3 T; however, finite magnetization develops above 3 T, the same range of magnetic fields where the resistivity peak develops (see Fig. \ref{fig:XX40_Rxx_map}). We thus associate the resistivity peak with the formation of an isospin polarized state at finite magnetic field.  This is consistent with the hypothesis that the anomalous resistivity peak that develops in in-plane magnetic field indeed marks the boundary between a polarized and an unpolarized phase, similar to the behavior in out-of-plane magnetic field in this density regime\cite{lu_superconductors_2019,saito_hofstadter_2020,wu_chern_2020,nuckolls_strongly_2020}. 



\begin{figure*}[ht!]
\includegraphics[width= 135mm]{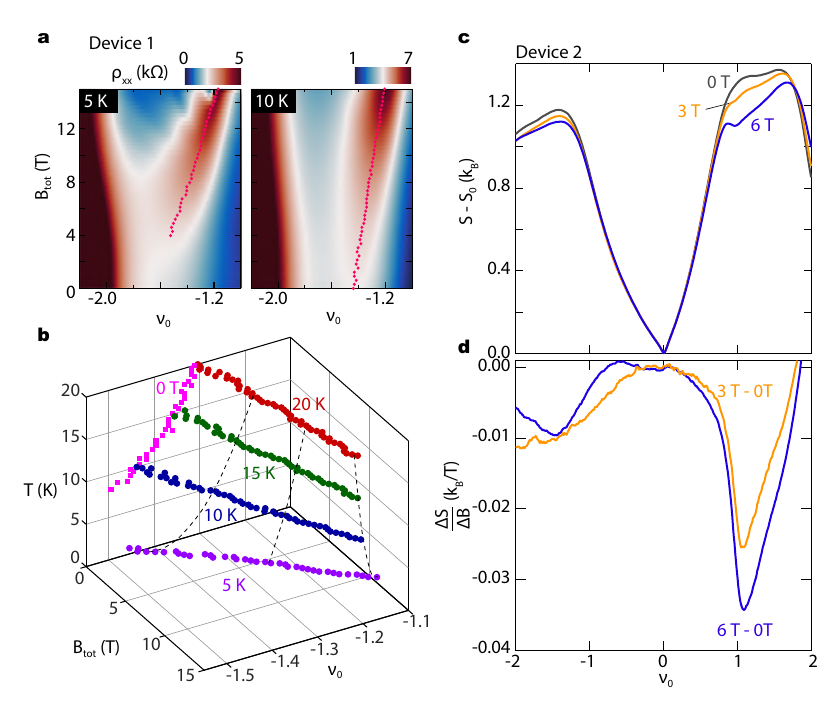}
 \caption{\textbf{Isospin Pomeranchuk effect and spin entropy.}
 \textbf{a}, $\rho_\mathrm{xx}$ as a function of $\nu_0$ and $B_\mathrm{tot}$ oriented at an angle of $9.1^\circ$ relative to the sample plane, measured at 5 and 10K in Device 1. Pink circles correspond to $\rho_\mathrm{xx}$ peak positions. 
\textbf{b}, The position of resistive peaks $\nu_\mathrm{peak}$ as a function of $B_\mathrm{tot}$ and $T$. The dashed curves are polynomial fits to the dots at $B_\parallel$ = 5, 10, 15T.
\textbf{c}, Entropy per superlattice unit cell measured relative to the charge neutrality point, $S-S_0$.  Entropy is derived from the finite difference between $\mu(\nu_0$ data measured at 4.2 and 12 K for $B_\parallel = 0, 3, 6$ T. 
\textbf{d}, $\frac{\Delta S}{\Delta B}$ as a function of $\nu_0$ at $T$ = 8.1 K, calculated from (S(6 T)-S(0 T))/6 T (blue) and (S(3 T)-S(0 T))/3 T (yellow).
}
\label{fig:3}
\end{figure*}

The apparent duality between $B_\parallel$- and $T$- dependent transport is suggestive of an entropically driven transition at finite temperature. In this scenario, the unpolarized Fermi liquid state has lower ground state energy than the IF$_3$ phase near $\nu=-1$, but the fluctuating moments of the IF$_3$ state make it entropically favorable at elevated temperature.  
The characteristic temperature scale at which these moments begin to fluctuate strongly, giving rise to a large isospin entropy, is given by the stiffness of the collective excitations of the spin, valley, and carbon sublattice degrees of freedom~\cite{bultinck_ground_2019}, which numerical calculations find to be in the few meV range\cite{khalaf_charged_2020,wu_collective_2020,khalaf_soft_2020,kumar_lattice_2020}. Combined with the expectation that ground state energies differ by similar energy scales, an entropically driven transition in the $\sim 10$ K regime is highly plausible. This is analogous to the well-known Pomeranchuk effect in $^3$He, where the liquid transforms into a solid upon raising the temperature. In our system, the role of the liquid phase is played by the unpolarized Fermi liquid, while the high temperature `solid' analog is the high temperature extension of the IF$_3$ phase, which, though it may have only negligible net magnetization is distinguished by the presence of local, strongly fluctuating magnetic moments.  

The connection between low $T$, high $B_\parallel$ and high $T$, $B_\parallel=0$ phases is confirmed by variable temperature measurements of $\rho_\mathrm{xx}$, shown in Fig. \ref{fig:3}a  where we plot $\rho_\mathrm{xx}$ as a function of $\nu_0$ and $B_\mathrm{tot}$ oriented at an angle of $9.1^\circ$ relative to the sample plane, measured at 5 and 10 K.
At the higher temperatures, the resistivity peak separating the high temperature extensions of the IU and IF$_3$ phases is visible at $B_\mathrm{tot}=0$ T, and is observed to move towards neutrality as a function of $B_\mathrm{tot}$ (see also Fig. \ref{linecut_highT})---precisely the expected behavior if the two high temperature phases differ in their spin polarization or magnetic susceptibilities.  
As shown in Figure \ref{fig:3}b, the resistivity peak can be used to map the boundary between the isospin symmetric phase prevailing at lower $B, T$, and $|\nu_0|$ and a state of finite spin susceptibility at higher temperatures, as shown in Figure \ref{fig:3}b. We note that a similar behavior of the phase boundary is both expected and observed near $\nu_0=+1$, even when the ground state is the IF$_3$ phase as in Device 2 (Fig. \ref{fig:XX40_peak}). 

In an out-of-plane $B_\perp$ the low-temperature magnetic transition appears to be first order for at least some range of $\nu$\cite{lu_superconductors_2019,saito_hofstadter_2020}, showing sharp jumps in experimental observables and hysteretic behavior. 
It is thus tempting to analyze the transitions in both in-plane $B_\parallel$ and as a function of $T$ in a first order framework as occurs in $^3$He.  
In this picture, the Zeeman energy difference, $E_Z^*=\mu_B B^*$, required to drive the transition in the $T\rightarrow0$ limit measures the ground state energy difference between the IU and IF$_3$ phases (or a strongly fluctuating version of the latter).  The temperature at which the transition occurs at $B=0$---which we by denote $T^*$---is determined by the condition that the entropy of the phase with fluctuating isospin moments overcomes this energy difference. One then expects an entropy jump of $\Delta S\approx E_z^*/T^*$ across the transition.  Using the peak position of resistivity to empirically define the transition, this estimate gives $\Delta S\approx .5\times k_B$ per superlattice unit cell (see Fig. \ref{S_transport}).

 A key prediction of the first order scenario is a jump in the entropy between the IU and IF$_3$ phases arising from disordering of isospins at finite temperature. We measured the total electronic entropy from the response of $\mu$ to changes in $T$ via the Maxwell relation $\left(\partial \mu/\partial T\right)_\nu=-\left(\partial S/\partial \nu\right)_T$, approximating $\partial \mu/\partial T$ from a finite difference between $\mu$ measurents at 4.2 K and 12 K.  
 Fig. \ref{fig:3}c shows the experimentally determined entropy measured relative to the entropy at the charge neutrality point, $S(\nu_0)-S_0$.  The entropy rises upon both electron or hole doping, reaching $\Delta S/k_B\sim1$ per superlattice unit cell near $\nu_0=\pm1$, where it levels off or even decreases.  However, we observed no jump, effectively ruling out a first order transition.  This is consistent with the absence of sharp features in either temperature- or $B_\parallel$-dependent transport measurements, suggesting the transition is either of higher order or simply a crossover.

 \begin{figure}[ht!]
\centering
\includegraphics[width=85mm]{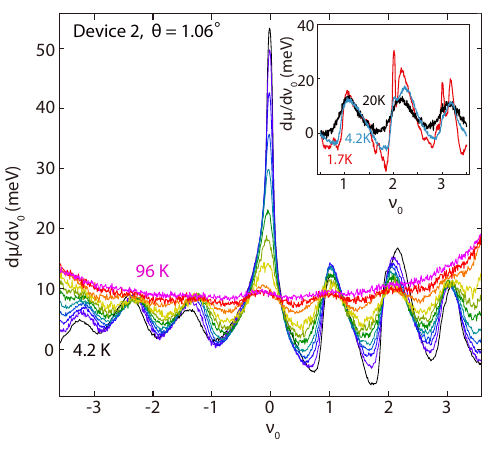}
 \caption{\textbf{
 Temperature dependence of the  inverse electronic compressibility $d\mu/d\nu_0$} The data is acquired at 4.2, 8, 12, 16, 20, 26, 32, 40, 59, 76 and 96 K.  Approximately commensurate oscillations appear at $T\lesssim 100$ K, the approximate scale of the Coulomb interactions. Thermodynamic gaps at $\nu_0=2,3$ appear only below 4.2 K, as shown in the inset which shows $d\mu/d\nu_0$ traces for 20 K, 4.2 K, and 1.7 K in black, cyan, and red respectively.
 }
\label{fig:4}
\end{figure}

Nevertheless, the existence of disordered isospins in the high temperature IF$_3$ phases is supported by the measured behavior of the entropy as a function of in-plane magnetic field. As shown in Fig. \ref{fig:3}d, $\Delta S$ decreases as a function of $B_\parallel$ for $\nu_0\gtrsim1$ and $\nu_0\lesssim -1$ corresponding to the high temperature IF$_3$ phases, but gives no experimentally detectable change in the IU phase.  
At temperatures of order the spin stiffness, the IF$_3$ phase is strongly fluctuating, leading to a spin dependent $S\approx k_B \ln2$ contribution to the entropy.  
This entropy can be suppressed by a Zeeman energy $E_Z\sim k_B T$.  We thus expect $\Delta S/\Delta B\sim \mu_B/T\approx0.08 k_B/\mathrm{Tesla}$ for the $T$ = 8 K temperature of our measurement.  We indeed observe an entropy suppression of this scale for $\nu_0\gtrsim1$, suggesting that the spin stiffness is indeed small and supporting the picture of a spin-entropy driven Pomeranchuk effect.  A smaller entropy suppression is observed for $\nu_0\lesssim-1$.  This could arise from a larger spin stiffness in the IF$_3$ phase at hole doping. The discrepancy between electron and hole doping highlights the quantitative importance of the particle-hole asymmetry of the underlying single particle wave functions, a problem 
recent theoretical literature has only begun to address\cite{kumar_lattice_2020,wu_collective_2020,khalaf_charged_2020,khalaf_soft_2020,bernevig_TBG_2020,vafek_towards_2020}. 


Our observation of an entropically driven transition suggests that soft neutral excitations of the electron system are likely play a key role in the physics of flat band moir\'e systems. While long-range magnetic order may appear only at low temperature or in the presence of a magnetic field, much of the phase diagram is dominated by the presence of large, strongly fluctuating local moments. A measurement of the compressibility $\partial \mu/\partial \nu_0$ as a function of density for a range of temperatures between 4.2 K and 96 K (Fig. \ref{fig:4}) shows a sequence of nearly commensurate, asymmetric peaks (see also Fig. \ref{Device2_highT}). These peaks have been interpreted\cite{zondiner_cascade_2020,wong_cascade_2020,park_flavour_2020} as indicating Fermi surface reconstruction due to a cascade of isospin symmetry breaking transitions.  Strikingly, however, our measurements show that the peaks in $\partial \mu/\partial \nu_0$ survive even at temperatures well above the scale of the spin stiffness, where no magnetic order is found.  
The compressibility features themselves disappear only at $T\approx 100$ K, comparable to the scale of the Coulomb interaction~\cite{andrei_graphene_2020}.  
We emphasize that in our thermodynamic measurements, the developement of finite magnetization is a detectable but subtle effect that does not qualitatively impact the the structure of the chemical potential (see Fig. \ref{chemical_potential}). It therefore seems likely that the peaks in $\partial\mu/\partial \nu_0$ mark the formation of local isospin moments, correlated only on length scales comparable to the moir\'{e} wavelength. 

The presence of strongly fluctuating isospin moments in much of the phase diagram should have a profound effect on the physics of tBLG. The small stiffness implied by our measurements may serve as an upper bound on superconducting $T_c$ in regions of the phase diagram where isospin order is observed, either because the isospin fluctuations act as pair breakers, or because the pairing mechanism itself requires isospin order. For example, isospin ordering is a prerequisite for the existence of skyrmion textures recently proposed to play a role in the superconductivity observed in tBLG\cite{khalaf_charged_2020}. 


Thermal disordering of the internal degrees of freedom is also expected to scatter electrons strongly at these temperatures.  Some portion, if not the majority, of the large high-temperature resistivity in flat band moir\'e systems likely arises from such scattering. This appears consistent with the experimentally observed ubiquity of both ferromagnetism across the flat band\cite{zondiner_cascade_2020,wong_cascade_2020} and large resistivity at intermediate temperatures\cite{polshyn_large_2019,cao_strange_2020}.  In addition, the `superconducting-like' transition observed in many moir\'e systems, in which the resistivity rises rapidly at temperatures of a few Kelvin, likely indicated the onset (at high temperatures) of this fluctuation moment phase.  The precise temperature dependence of the resistivity is not expected to be universal, depending on the details of the collective excitations and their coupling to the itinerant conduction electrons.  This is consistent with experimental observation of strong $\nu_0$-dependence of $\rho_\mathrm{xx}(T)$ (see  Fig. \ref{RT}).

\widetext
\let\oldaddcontentsline\addcontentsline
\renewcommand{\addcontentsline}[3]{}
\section*{Methods}
\noindent\textbf{Device fabrication and transport measurements}\\ In this study, we used two tBLG devices; Device 1 (1.12$^\circ$) and 2 (1.06$^\circ$). Both devices were fabricated using a ``cut-and-stack'' technique described in Ref. \onlinecite{saito_independent_2020}. Device 1 is the same as device \#5 in Ref. \onlinecite{saito_independent_2020} and the device studied in Ref. \onlinecite{saito_hofstadter_2020}. Device 2 is the same as the device used in Ref. \onlinecite{liu_tuning_2020}. Prior to stacking, we first cut graphene into two pieces using AFM to prevent the unintentional strain in tearing graphene. We used a poly(bisphenol A carbonate) (PC)/polydimethylsiloxane (PDMS) stamp mounted on a glass slide for stacking tBLG heterostructures. The final structure of Device 1 and 2 are hBN(40 nm)-tBLG-hBN(40 nm)-graphite and graphite-hBN(30 nm)-BLG-hBN(3 nm)-tBLG-hBN(30 nm)-graphite as shown in Fig. \ref{devices}.
Electrical connections to the tBLG were made by CHF$_3$/O$_3$ etching and deposition of the Cr/Pd/Au (2/15/180 nm) metal edge-contacts for Device 1 and Cr/Au (2/100 nm) metal edge-contacts for Device 2 \cite{wang_one-dimensional_2013}. 

Transport data in Figure 1a-c and d (0.5, 2.5, 4.5 K) were acquired with Device 1 in a top-loading cryogen-free dilution refrigerator with a nominal base temperature of 10 mK, using a probe with heavy RF filtering at an excitation current of 2 nA at a frequency of 17.777 Hz.  
Data in Figures 2a-c were acquired in a different probe without filtering at an excitation current of 10 nA at a frequency of 278 Hz.  
Data in Figures 1d (8 and 20 K), 2d and 3a were acquired using a wet, sample-in-vapor variable temperature system without filtering at an excitation current of 10 nA at a frequency of 278 Hz.

We measure the geometric capacitance in Device 1 from the Hall density near the charge neutrality point to be be $c_g=58.2 \pm 0.3$ nF/cm$^2$. Separate measurements using quantum oscillations give comparable results, $c_g= 58.5 \pm 0.1$ nF/cm$^2$.  
The twist angle $\theta$ is determined from the values of charge carrier density at which the insulating states at $n_{\nu_0 = \pm 2}$ are observed, following $n_{\nu_0 = \pm 2} = \pm 4 \theta^2/\sqrt{3}a^2$ , where $a$= 0.246 nm is the lattice constant of graphene. From these measurements, $\nu_0=A_\mathrm{u.c} c_g/e (v_g-v_g^0)$ where $v_g^0$ is the gate voltage corresponding to charge neutrality and $A_\mathrm{u.c.} c_g/e=0.5026$ V$^{-1}$. 

\vspace{5pt}

\noindent\textbf{Thermodynamic determination of $\mu$, $M$, and $S$}\\
The entropy ($S$) and the magnetization ($M$) are determined by measuring temperature dependent and in-plane magnetic field dependent chemical potential $\mu$ and making use of the Maxwell relations 
\begin{align}
    \left(\frac{\partial{S}}{\partial{\nu}} \right)_{T} & = -\left(\frac{\partial{\mu}}{\partial{T}} \right)_{\nu} & 
    \left(\frac{\partial{M}}{\partial{\nu}} \right)_{B} & = -\left(\frac{\partial{\mu}}{\partial{B}} \right)_{\nu}
    \label{dmu}
\end{align}
where $\nu$ is the filling factor. 
$S$ and $M$ can then be determined by integrating the measured right hand sides of Eq. \ref{dmu} with respect to $\nu$.
We perform the thermodynamic measurements on Device 2, which consists of a $1.06^\circ$ angle tBLG separated by a 3 nm BN spacer from a Bernal bilayer flake that is separately contacted (Fig. \ref{devices}b). Transport data from this device was described in \cite{liu_tuning_2020}.  To determine the chemical potential,  we use the measurement technique described in Ref.\cite{yang_experimental_2020}. In this technique, an excitation current (5 nA-50 nA) with frequency $f_1=321$ Hz is used to measure the four-terminal resistance $R_\mathrm{BLG}$ of the Bernal bilayer graphene, while a second frequency $f_2=123$ Hz is used to modulate the top gate voltage resulting in a measurable desponse at frequency $f_1-f_2$ proportional to $dR_\mathrm{BLG}/dV_{tg}$; crucially, this response vanishes at a resistivity extremum such as the charge neutrality point.  A feedback loop is then used to maintain $dR_\mathrm{BLG}/dV_\mathrm{tg}=0$ as the bottom gate is changed by applying a feedback voltage to the twisted bilayer.  The output voltage of this feeback loop is then equal to $\mu/e$. IN all measurements, the displacement field of the Bernal bilayer graphene is maintained at $D=14$ mV/nm.

Strictly speaking, our technique measures $\mu_\mathrm{tBLG}(\nu)-\mu^\mathrm{CNP}_\mathrm{bBLG}$, the chemical potential difference between the twisted bilayer and the charge-neutral bernal bilayer detector.  While the change in $\mu^\mathrm{CNP}_\mathrm{bBLG}$ with temperature and magnetic field is small, so are differences in $\mu_\mathrm{tBLG}(\nu)$. To fix the possible offset between curves measured under different conditions, we set $d\mu/dB_\parallel$ and $d\mu/dT$ to be zero at $\nu_0=0$.  These curves are then integrated from $\nu_0=0$ filling factors. We thus measure $S-S_0$ and $M_\parallel-M_{\parallel,0}$, the changes in $S_\parallel$ or $M_\parallel$ relative to their values at charge neutrality: 

\begin{align}
    S(\nu_0) -S_0 & = \int_0^{\nu_0}\left(\frac{\partial{S}}{\partial{\nu}}\right)d \nu -\nu_0\frac{\partial{S}}{\partial{\nu}}\bigg|_{\nu_0=0} &
    M_\parallel(\nu_0)-M_{\parallel,0} & = \int_0^{\nu_0}\left(\frac{\partial{M}}{\partial{\nu}}\right)d\nu-\nu_0 \frac{\partial{M}}{\partial{\nu}}\bigg|_{\nu_0=0}
\end{align}

In the case of $M_\parallel$, we expect $M_\parallel$ due to the absence of in-plane B dependence of measured quantities as well as from the flat behavior of $dM_\parallel/d\nu$ in that region.  
However, in the case of $S$, charge neutrality may well develop a large entropy both from excited quasiparticles at high temperature or from modes associated with breaking of additional symmetries in the charge neutral state.  However, we note that for the purposes of correlating the behavior of transport near $\nu=\pm1$ with the state surrounding charge neutrality, $S-S_0$ is the relevant quantity.


\vspace{5pt}

\noindent\textbf{Thermodynamic model}\\

To describe the phase transition between the IU to the IF$_{3}$ phases,
we write their free energy per moir\'{e} unit cell as:

\begin{equation}
f_{i}(\nu,T)=e_{i}+\frac{1}{2}\left(\frac{1}{\kappa_{i}}+\frac{e^{2}}{c_{g}}\right)\nu^{2}+\mu_{i}\nu-\frac{1}{2}\gamma_{i}T^{2}-s_{i}T,
\end{equation}
where $i=1,2$ corresponds to the IU and IF$_{3}$ phases, respectively,
$e_{i}$ is an offset energy, $c_{g}$ is the geometric capacitance
to the gate per moir\'{e} unit cell, $\mu_{i}$ is an offset chemical potential, $\gamma_{i}$
is the specific heat coefficient (both phases are assumed to be metallic, despite the fact that the IF$_3$ phase has large, fluctuating magnetic moments),
and $\kappa_{i}$ is the compressibility (or quantum capacitance).
$s_{i}$ is a temperature-independent contribution of the entropy.
The IU phase is a Fermi liquid whose entropy is proportional to temperature, hence $s_{1}=0$. In the IF$_{3}$
phase, the fluctuating moments give a contribution $s_{2}>0$ to the
entropy at temperatures exceeding spin stiffness.

Since the experiment is carried out at a constant gate voltage, the phase
transition (assumed to be of first order in the absence of disorder)
occurs when the Landau grand potential $\Omega_{i}(v_{g},T)=f_{i}(\nu,T)-ev_{g}\nu$
of the two phases are equal. We minimize the grand potentials of each phase with respect
to $\nu$, and express the grand potentials in terms of the reference
filling factor $\nu_{0}\equiv\frac{1}{e}c_{g}v_{g}$. The transition
line in the $(\nu_{0},T)$ plane is then given by the condition:
\begin{equation}
\Omega_{2}-\Omega_{1}=\Delta \tilde{e}(\nu_{0})-\frac{1}{2} \Delta \gamma T^{2}- T \Delta s =0,
\end{equation}
where $\Delta \tilde{e}(\nu_{0})\approx e_{2} - e_1 -\frac{1}{2}\left(\frac{1}{\kappa_{2}}-\frac{1}{\kappa_{1}}\right)\nu_{0}^{2} + (\mu_{2} - \mu_{1})\nu_{0}$, $\Delta \gamma = \gamma_2 - \gamma_1$, and $\Delta s = s_2 - s_1$. 
Here, we have used the fact that in our setup $\frac{e^{2}}{c_{g}}\gg\frac{1}{\kappa_{1,2}}$,
and neglected terms that are suppressed by factors of $\frac{c_{g}}{e^{2}\kappa_{1,2}}$.
At sufficiently low temperature compared to the bandwidth (estimated to be of the order of $200-300$K), the quadratic term in $T$ is much
smaller than the linear term, giving a transition at 
\begin{equation}
T^{*}=\frac{\Delta \tilde{e}(\nu_{0})}{\Delta s}.
\label{eq:Tstar}
\end{equation}

To determine the entropy $\Delta s$ from the experiment, we need an estimate of $\Delta e(\nu_0)$. This can be obtained by examining the magnetic field needed to trigger the transition from the IU to the IF$_3$ phase at low temperature (below the spin stiffness in the IF$_3$ phase). We consider an in-plane field, assuming that it acts primarily through the Zeeman effect. The magnetic field induces an additional term in
the grand potentials equal to $-\int_0^B m_{i}(B')dB'$, where $m_{i}(B)$ is
the magnetic moment per moir\'{e} unit cell. At sufficiently low temperature,
where the excess magnetic entropy $s_{2}$ of the IF$_{3}$ phase is
quenched, this phase is spin polarized, and its magnetization is nearly field-independent. To  the magnetic moment
in this phase, we assume that one isospin flavor whose spin is antiparallel to the
Zeeman field is completely empty (i.e., this flavor has a filling of one hole
away from charge neutrality), whereas the other three flavors are
equally populated~\cite{zondiner_cascade_2020}. These considerations give a magnetic moment of
$m_{2}=\mu_{B}\frac{4+\nu_{0}}{3}$ in the IF$_{3}$ phase. In
contrast, the IU phase has no magnetic moment at $B=0$. Since the IU phase is a Fermi liquid, its magnetization is proportional to the ratio between the Zeeman energy and the bandwidth, and is much smaller than that of the IF$_3$ phase. We therefore neglect the magnetization of the IU
phase, $m_1\approx 0$. The field-driven transition at low temperature occurs when
\begin{equation}
\Delta \tilde{e}(\nu_{0})-\mu_{B}\frac{4+\nu_0}{3}B^{*}=0.
\label{eq:Bstar}
\end{equation}
Combining Eqs.~(\ref{eq:Tstar},\ref{eq:Bstar}) gives 
\begin{equation}
    \Delta s = \frac{E^*_Z(\nu_0)}{T^*(\nu_0)}
    \frac{4+\nu_0}{3}
    \label{dS}
\end{equation}
We plot the expected entropy in Fig. \ref{S_transport}.

\let\addcontentsline\oldaddcontentsline
\let\oldaddcontentsline\addcontentsline
\renewcommand{\addcontentsline}[3]{}

\section*{Acknowledgments}
\vspace{-12pt}
\noindent
The authors acknowledge discussions with S. Kivelson, A. Macdonald, B. Spivak, and M. Zaletel, as well as experimental assistance from H. Polshyn and C. Tshirhart. Y.S. acknowledges the support of the Elings Prize Fellowship from the California NanoSystems Institute at University of California, Santa Barbara.
K.W. and T.T. acknowledge support from the Elemental Strategy Initiative conducted by the MEXT, Japan and the CREST (JPMJCR15F3), JST.
Transport and fabrication experiments at UCSB were supported by the ARO under MURI W911NF-16-1-0361. Thermodynamic measurements were supported by the National Science Foundation under \# DMR-1654186. 
A.F.Y. acknowledges the support of the David and Lucille Packard Foundation under award 2016-65145. EB was supported by the European Research Council (ERC) under grant HQMAT (grant no. 817799), and by the US-Israel Binational Science Foundation (BSF) under the NSF-BSF DMR program (grant no. \# DMR-1608055). 
X. L. and J.I.A.L are supported by Brown University.
\let\addcontentsline\oldaddcontentsline
\let\oldaddcontentsline\addcontentsline
\renewcommand{\addcontentsline}[3]{}

\section*{Author Contributions}
\vspace{-12pt}
\noindent
Y.S., J.G. and X.L. fabricated tBLG devices. Y.S. and F.Y. performed the measurements. Y.S., F.Y. and A.F.Y analyzed the data. A.F.Y. and E.B. constructed the thermodynamic model. 
Y.S., F.Y., E.B. and A.F.Y wrote the paper with inputs from J.I.A.L. 
T.T. and K.W. grew the hBN crystals.

\let\addcontentsline\oldaddcontentsline

\let\oldaddcontentsline\addcontentsline
\renewcommand{\addcontentsline}[3]{}
\section*{Competing interests}
\vspace{-12pt}
\noindent
The authors declare no competing financial interests.
\let\addcontentsline\oldaddcontentsline

\let\oldaddcontentsline\addcontentsline
\renewcommand{\addcontentsline}[3]{}
\bibliographystyle{unsrt}
\bibliography{references}
\let\addcontentsline\oldaddcontentsline

\clearpage

\pagebreak

\begin{center}
\textbf{\large Supplementary Information for \\Isospin Pomeranchuk effect and the entropy of collective excitations in twisted bilayer graphene}\\
\vspace{10pt}
Yu Saito, Fangyuan Yang, Xiaoxue Liu, Jingyuan Ge,  Kenji Watanabe, Takashi Taniguchi, J.I.A. Li, Erez Berg and Andrea F. Young\\
\vspace{10pt}
\end{center}
\renewcommand{\thefigure}{S\arabic{figure}}
\renewcommand{\thesubsection}{S\arabic{subsection}}
\setcounter{secnumdepth}{2}
\renewcommand{\theequation}{S\arabic{equation}}
\renewcommand{\thetable}{S\arabic{table}}
\setcounter{figure}{0}
\setcounter{equation}{0}
\onecolumngrid

\begin{figure*}[ht!]
\includegraphics[width= 7.2in]{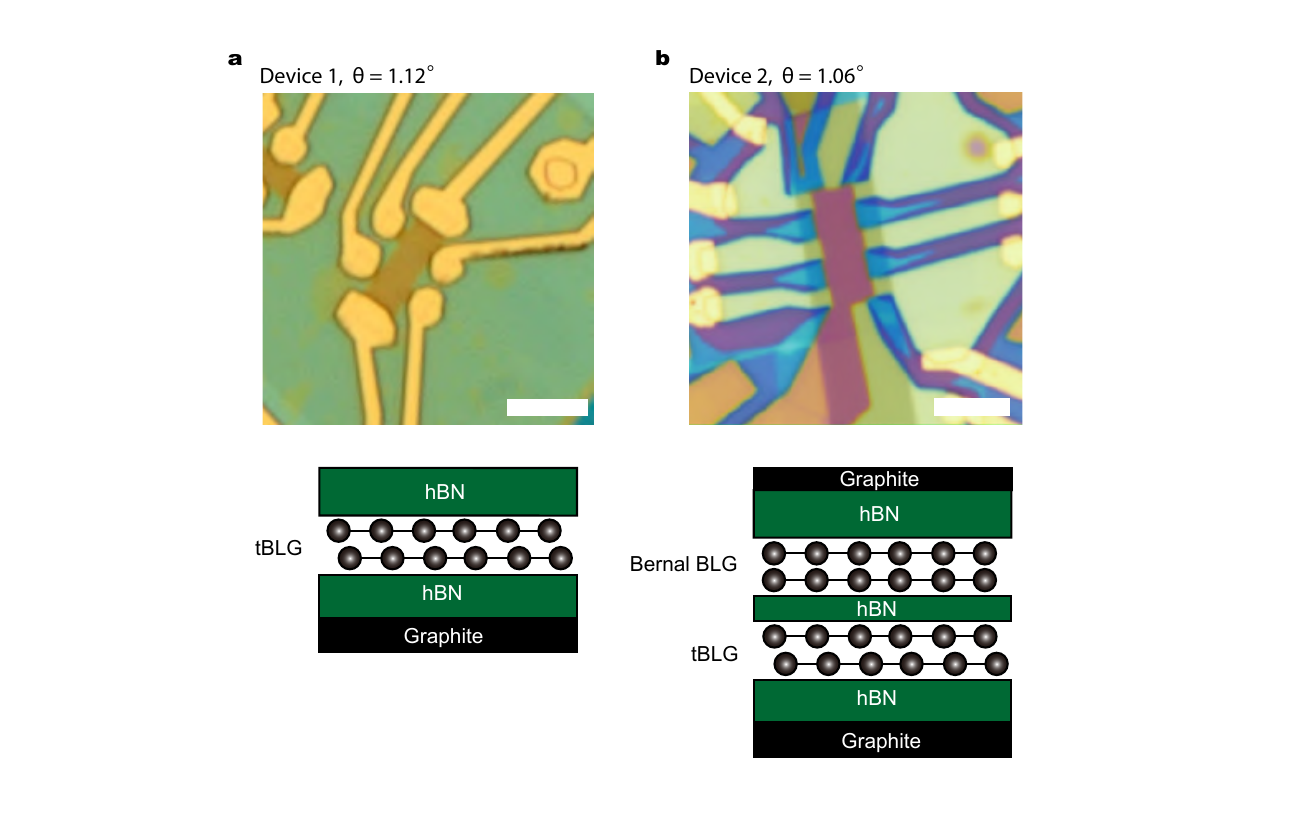}
\caption{\textbf{tBLG devices}
\textbf{a}, Optical images of the Device 1 and 2. Scale bar corresponds to 5 $\mu$m. 
\textbf{b}, Schematic of tBLG heterostructures for Device 1 and 2.}.
\label{devices}
\end{figure*}

\begin{figure*}[ht!]
\includegraphics[width= 6.8in]{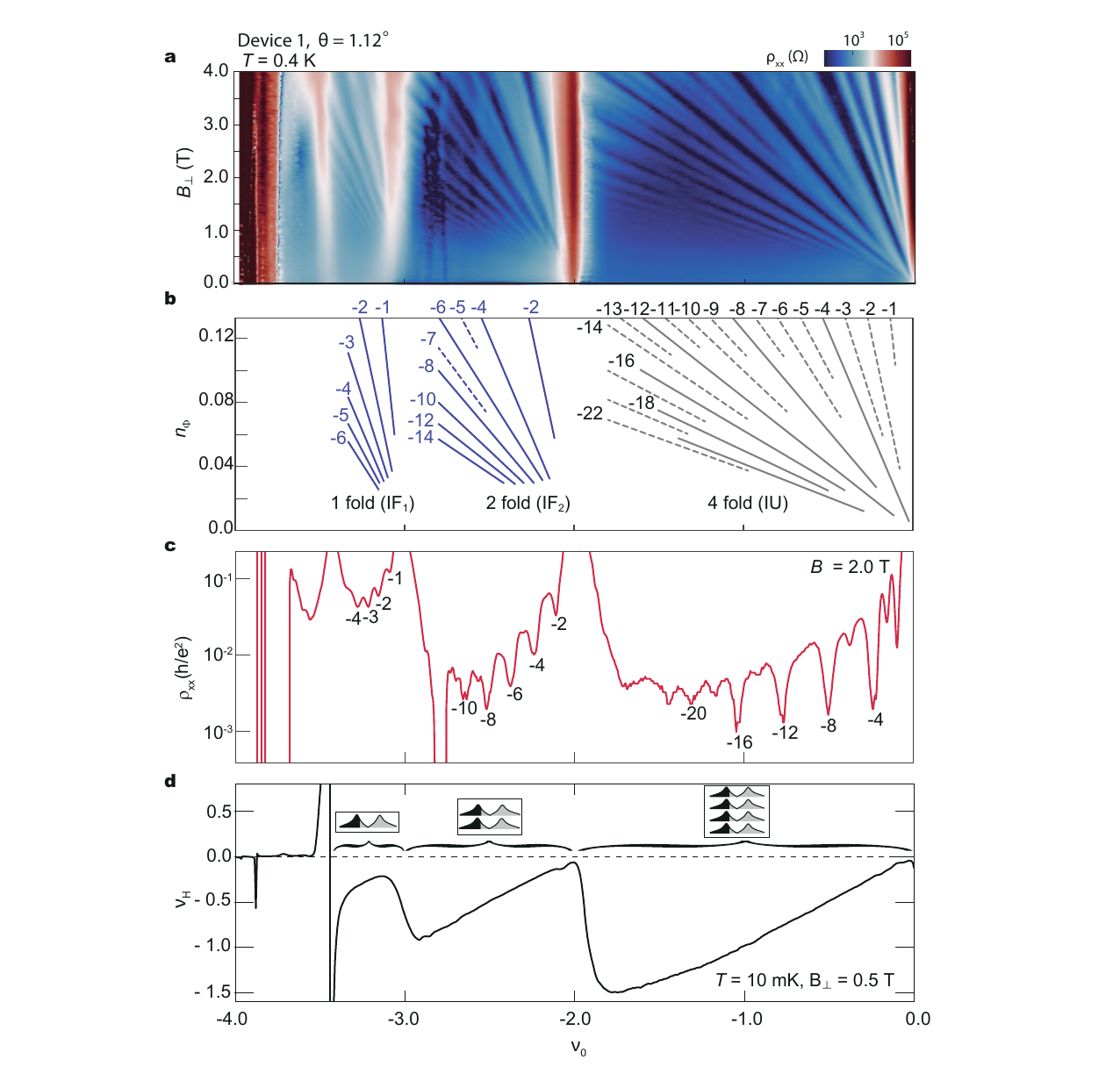}
\caption{\textbf{Landau fan diagram in Device 1.}
 \textbf{a}, $\rho_\mathrm{xx}$ as a function of $\nu_0$ and $B$ at 0.4 K.
 \textbf{b}, Schematic observed Chern insulator structure based on $\textbf{a}$.
 \textbf{c}, Line cuts of $\rho_\mathrm{xx}$ as a function of $\nu_0$ at 2.0 T in \textbf{a}.
 \textbf{d}, Hall density $\nu_H$ versus $\nu_0$ at $B_\perp = 0.5$ T and 20 mK. Inset: Schematic images show the pattern of flavor filling consistent with observed low magnetic field quantum oscillations \cite{saito_hofstadter_2020}}.
\label{fig:low field fan}
\end{figure*}

\begin{figure*}[ht!]
\centering
\includegraphics[width= 7.2 in]{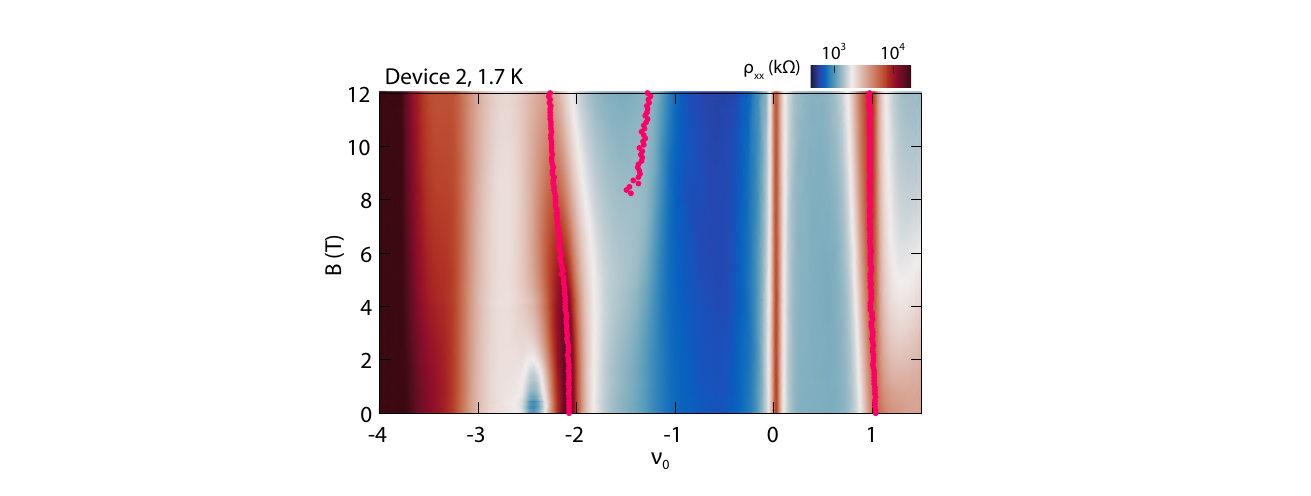}
 \caption{\textbf{$\rho_\mathrm{xx}$ a function of $B_\parallel$ and $\nu_0$ at $T = 1.7$ K in Device 2.} 
 Pink circles correspond to the position of the local maxima in  $\rho_\mathrm{xx}$.}
\label{fig:XX40_Rxx_map}
\end{figure*}

\begin{figure*}[ht!]
\includegraphics[width= 7.2in]{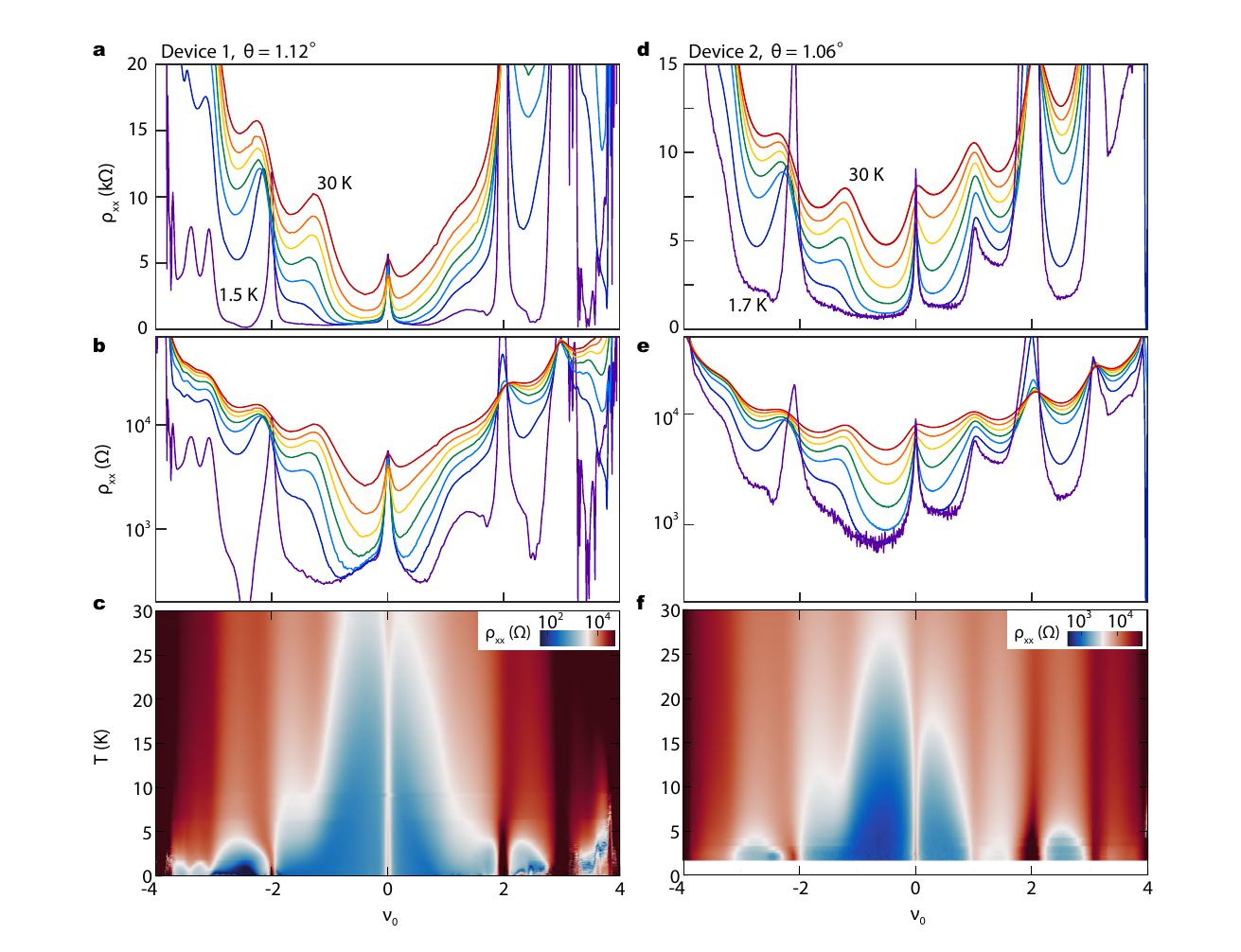}
\caption{\textbf{Temperature dependence of resistivity $\rho_\mathrm{xx}$}. 
\textbf{a,b,d,e} $\rho_\mathrm{xx}$ as a function of nominal filling factor $\nu_0$ at various temperatures up to 30 K in Device 1 (\textbf{a}:linear scale, \textbf{b}:log-scale) and Device 2 (\textbf{d}: linear-scale, \textbf{e}: log-scale). The traces in \textbf{a, b} are measured at $T$ = 1.5, 5, 8, 12, 17, 22, 30 K and the traces in \textbf{d, e} are measured at $T$ = 1.7, 5, 10, 15, 20, 25, 30 K.
\textbf{c, f,} 2D map of $\rho_\mathrm{xx}$ as a function of $\nu_0$ and $T$ in Device 1 (\textbf{c}) and 2 (\textbf{f}).}
\label{fig:Rxx_comp}
\end{figure*}

\begin{figure*}[ht!]
\includegraphics[width= 7.2in]{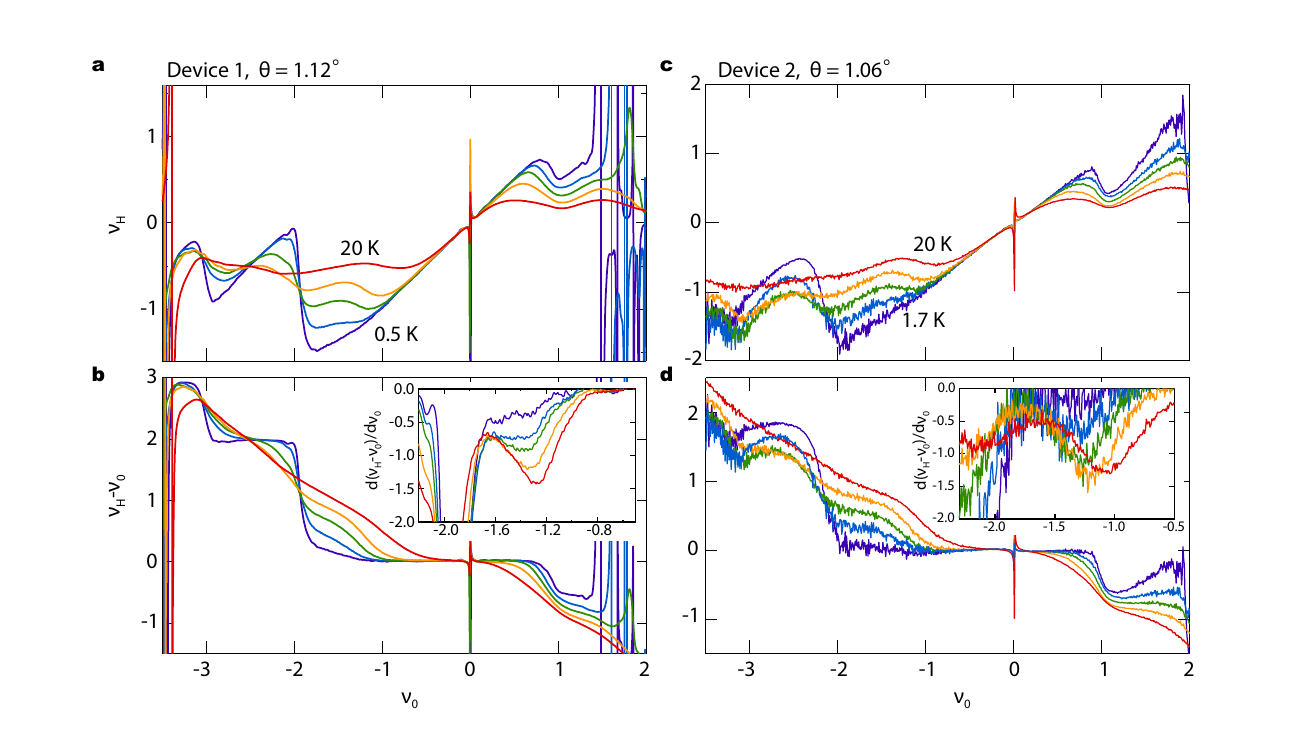}
\caption{\textbf{Temperature dependence of Hall density behavior.}
\textbf{a,c,} Hall density $\nu_\mathrm{H}$ expressed in electrons per superlattice unit cell as a function of $\nu_0$ up to 20 K at fixed $B_\mathrm{\perp}$ = 0.5 T. 
The data of Device 1 is measured with $T$ = 0.5, 2.5, 4.5, 8, 20 K (\textbf{a}) and the data of device 2 is measured with $T$ = 1.7, 4.3, 6, 10, 20 K (\textbf{c}).
\textbf{b,d,}  Subtracted Hall density $\nu_\mathrm{H}-\nu_0$ as a function of $\nu_0$ at each temperature 
in Device 1 (\textbf{b}) and 2 (\textbf{d}) Inset: $d(\nu_H-\nu_0)/d\nu_0$ as a function of $\nu_0$ at each temperature around $\nu_0 = -1$. $d(\nu_H-\nu_0)/d\nu_0$ is calculated from $\nu_\mathrm{H}-\nu_0$ using a 20-point moving average (\textbf{b}) and a 40-point moving average (\textbf{d}) in $\nu_0$.}.
\label{fig:vH_comp}
\end{figure*}


\begin{figure*}[ht!]
\includegraphics[width= 6.8in]{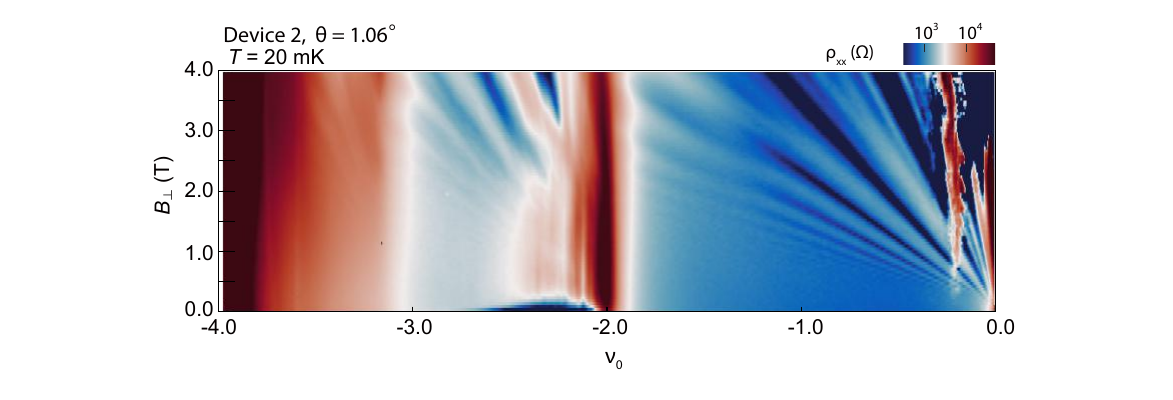}
\caption{\textbf{Landau fan diagram in Device 2.}
$\rho_\mathrm{xx}$ as a function of $\nu$ and $B$ at $T$ = 20 mK. The data is acquired when the displacement field at BLG is -200 mV/nm and density of BLG is 0.}
\label{fig:xx40_landaufan}
\end{figure*}

\begin{figure*}[ht!]
\centering
\includegraphics[width= 7.2 in]{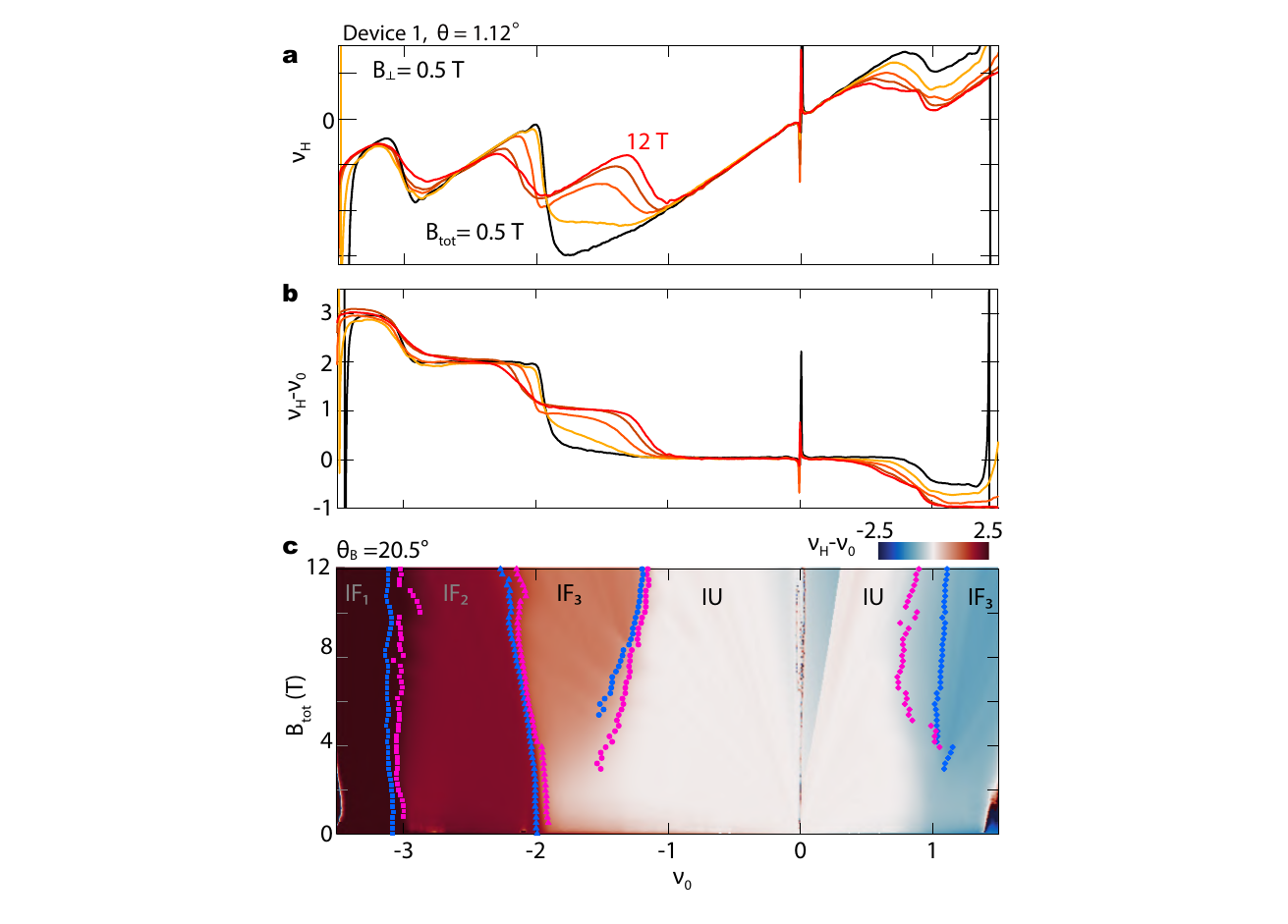}
 \caption{\textbf{In-plane magnetic field dependence of Hall density in Device 1.} \textbf{a}, Hall density $\nu_\mathrm{H}$ and \textbf{b} subtracted Hall density $\nu_\mathrm{H}-\nu_0$ expressed in electrons per superlattice unit cell, and measured with $B_\mathrm{tot}$ = 0.5, 3, 6, 9, 12 T and fixed $B_\mathrm{\perp}$ = 0.5 T.
\textbf{c}, $\nu_\mathrm{H}-\nu_0$ as a function of $B_\mathrm{tot}$ and $\nu_0$ with $\theta_\mathrm{B}$ = 20.5$^{\circ}$, measured at nominal $T$ = 20 mK. Blue and pink circles correspond to the positions of peaks of $\rho_\mathrm{xx}$ and the points of maximum descent in $\nu_\mathrm{H}-\nu_0$, respectively, and denote phase boundaries between symmetry breaking isospin ferromagnets (IF$_1$, IF$_2$, and IF$_3$) and an isospin unpolarized state (IU).}
\label{fig:vH}
\end{figure*}

\begin{figure*}[ht!]
\centering
\includegraphics[width= 6.8 in]{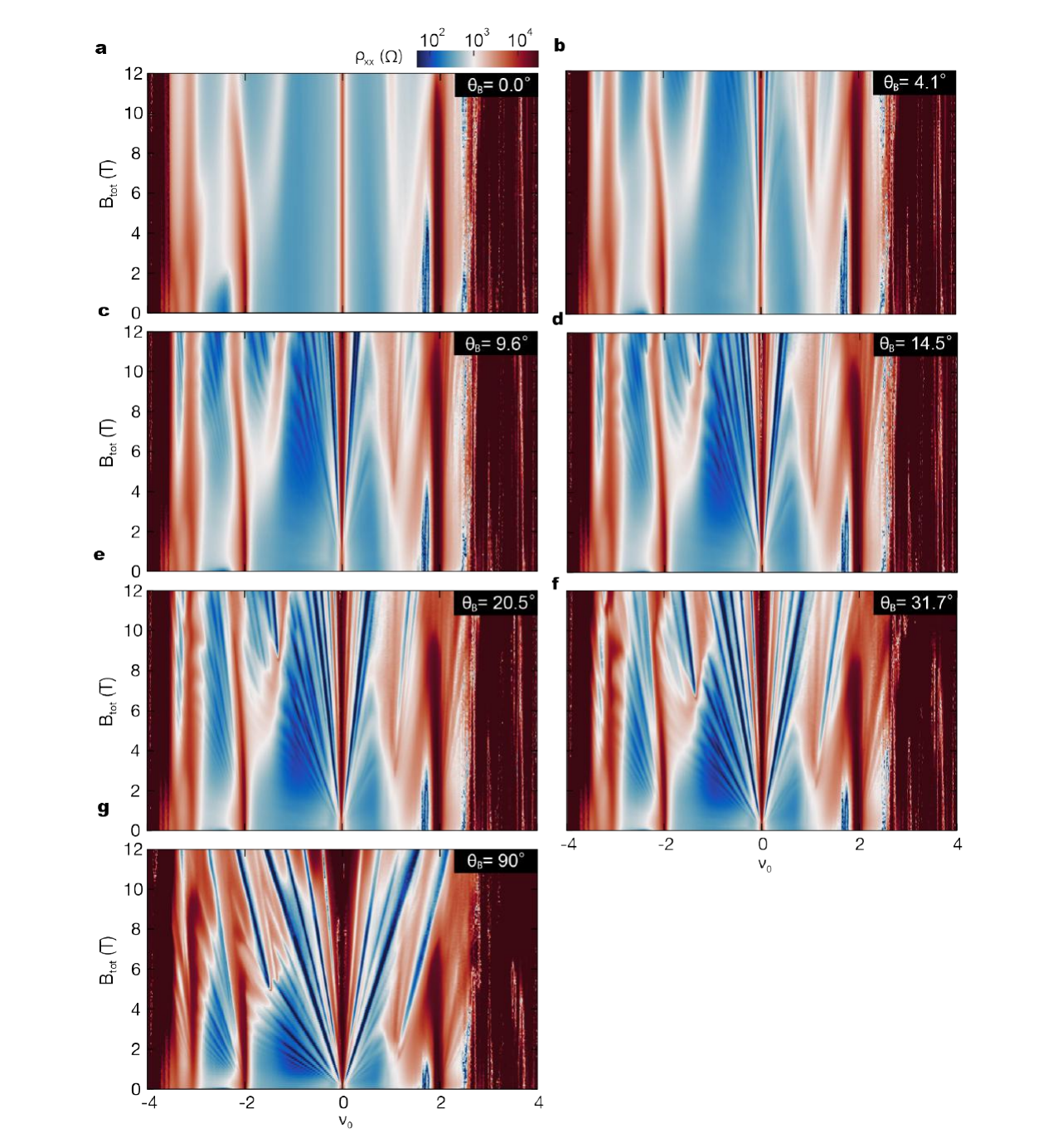}
 \caption{\textbf{Landau fan diagrams for different magnetic field angles at a nominal $T$ = 20 mK in Device 1.}
\textbf{a-g}, $\rho_\mathrm{xx}$ as a function of nominal filling factor $\nu_0$ and total magnetic field $B_\mathrm{tot}$ oriented at an angle with respect to the plane $\theta_\mathrm{B}$ of 0.0$^{\circ}$ (\textbf{a}), 4.1$^{\circ}$ (\textbf{b}), 9.6$^{\circ}$ (\textbf{c}), 14.5$^{\circ}$ (\textbf{d}), 20.5$^{\circ}$ (\textbf{e}), 31.7$^{\circ}$ (\textbf{f}) and 90.0$^{\circ}$ (\textbf{g}).}
\label{all_landaufans_rxx}
\end{figure*}

\begin{figure*}[ht!]
\centering
\includegraphics[width= 7.4 in]{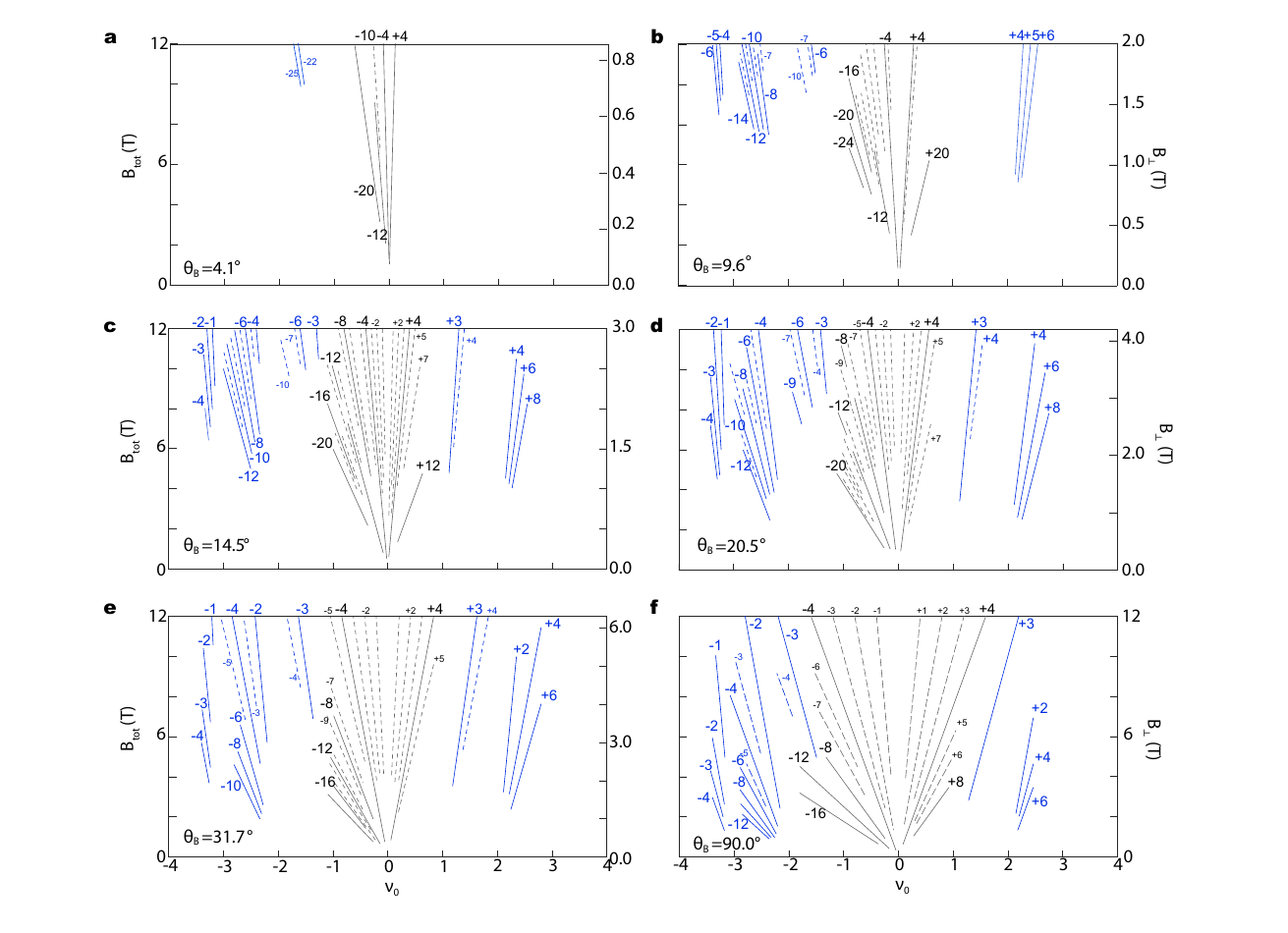}
 \caption{\textbf{Schematic Landau fan diagrams for different magnetic field angles.} \textbf{a-f}, The Schematic landau fan structures are based on \textbf{b-g} in Fig. \ref{all_landaufans_rxx}.
 }
\label{tiltedLL_analysis}
\end{figure*}

\begin{figure*}[ht!]
\includegraphics[width= 7.6in]{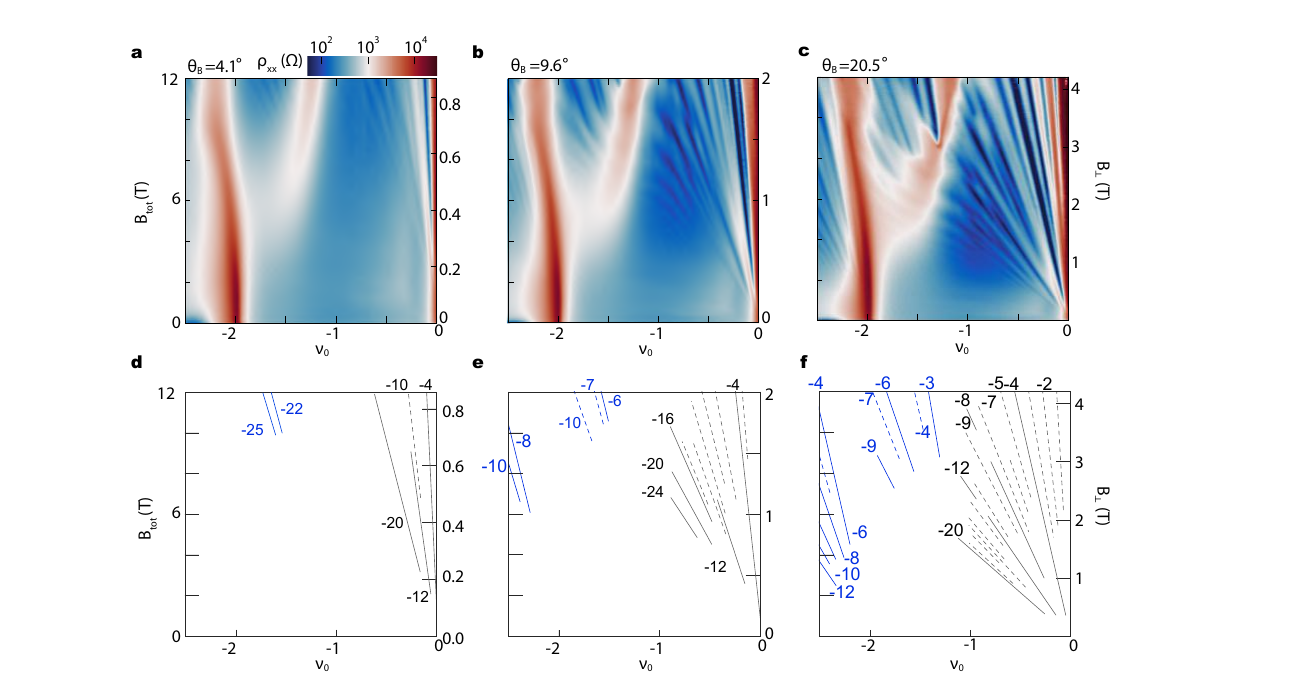}
\caption{\textbf{Landau fan diagram at hole side in tilted magnetic field in Device 1.}
\textbf{a}, $\rho_\mathrm{xx}$ as a function of  $\nu_0$ and total magnetic field $B_\mathrm{tot}$ oriented at an angle with respect to the plane $\theta_\mathrm{B}$ of 4.1$^{\circ}$(\textbf{a}), 9.6$^{\circ}$(\textbf{b}), 20.5$^{\circ}$(\textbf{c}).
\textbf{d-f}, Schematics of Landau fan diagram based on \textbf{a-c}.}
\label{fig:supple_LL_magni}
\end{figure*}

\begin{figure*}[ht!]
\includegraphics[width= 6.5in]{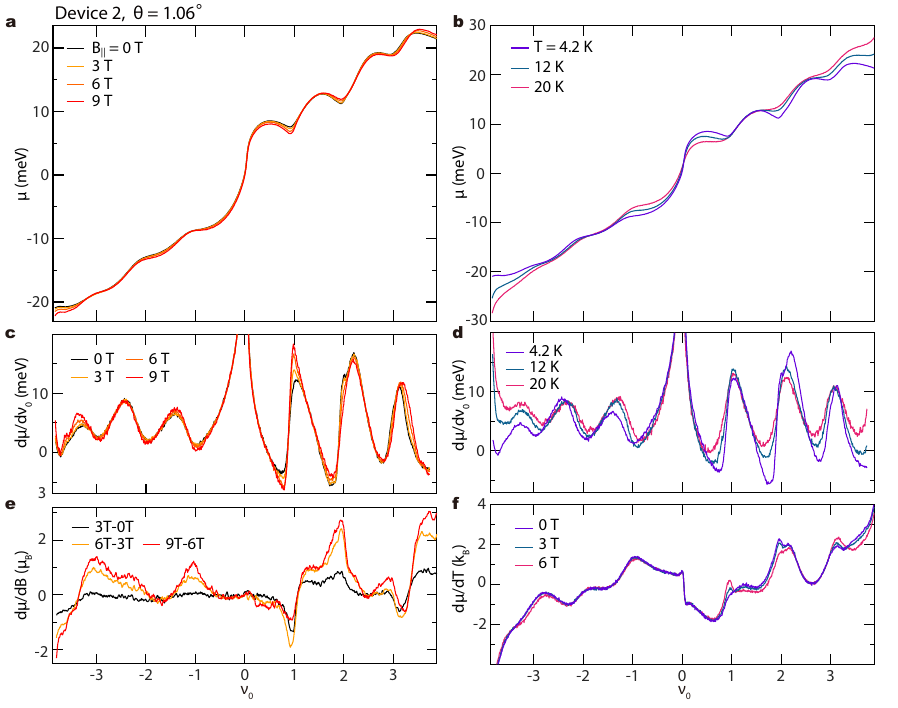}
\caption{\textbf{Thermodynamic measurements in Device 2.}
\textbf{a}, Chemical potential $\mu$ as a function of $\nu_0$ at $T$ = 4.2 K and $B_\parallel = $ 0, 3, 6, 9 T.
\textbf{b}, $\mu$ as a function of $\nu_0$ at $B$ = 0 T and $T = $ 4.2, 12, 20 K. 
\textbf{c}, inverse compressibility $d\mu/d\nu$ as a function of $\nu_0$ at T = 4.2 K and $B_\parallel = $ 0, 3, 6, 9 T. 
\textbf{d}, $d\mu/d\nu$ as a function of $\nu_0$ at $B$ = 0 T and $T = $ 4.2, 12, 20 K. 
\textbf{e}, $d\mu/dB_\parallel$ as a function of $\nu_0$ at $T$ = 4.2 K, calculated from ($\mu$(9 T)-$\mu$(6 T))/3 T, ($\mu$(6 T)-$\mu$(3 T))/3 T and ($\mu$(3 T)-$\mu$(0 T))/3 T. 
\textbf{f}, $d\mu/dT$ as a function of $\nu_0$, calculated from ($\mu$(12 K)-$\mu$(4.2 K))/7.8 K at $B_\parallel=$ 0, 3, 6 T}.
\label{chemical_potential}
\end{figure*}


\begin{figure*}[ht!]
\centering
\includegraphics[width= 7.2 in]{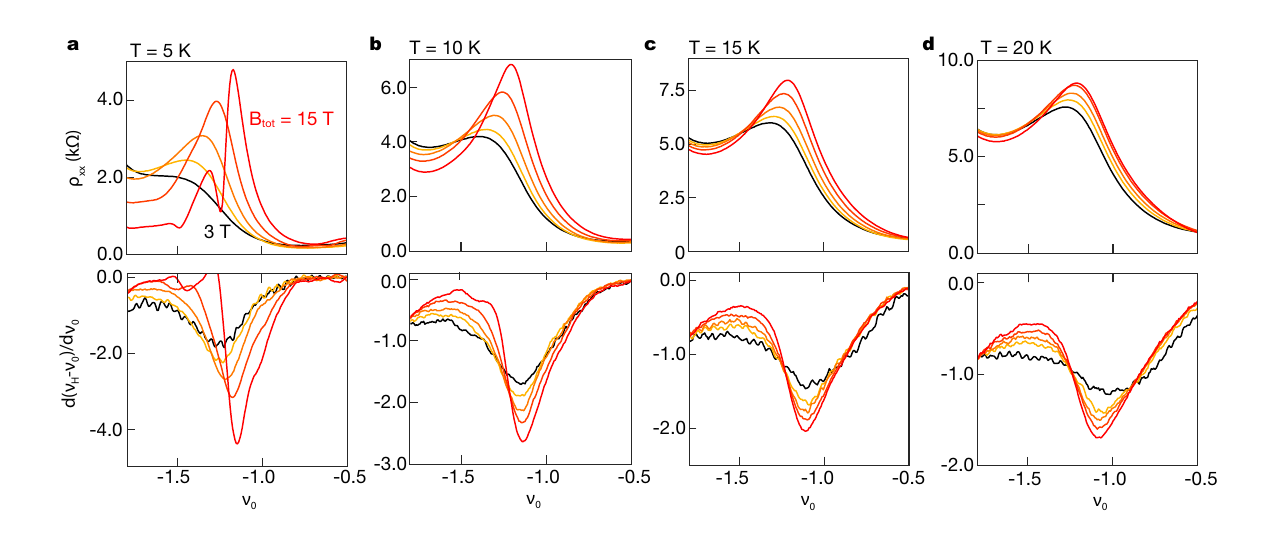}
 \caption{\textbf{High temperature transport in tilted magnetic field in Device 1.}
$\rho_\mathrm{xx}$ (top) and $d(\nu_\mathrm{H}-\nu_0)/d\nu_0$ (bottom) as a function of $\nu_0$ at $T$ = 5 (\textbf{a}), 10 (\textbf{b}), 15 (\textbf{c}) and 20 K (\textbf{d}) at $B_\mathrm{tot}$ = 3, 6, 9, 12, 15 T oriented at an angle of $9.1^\circ$ relative to the plane. 
$d(\nu_\mathrm{H}-\nu_0)/d\nu_0$ is calculated from $\nu_\mathrm{H}-\nu_0$ using a 20-point moving average in $\nu_0$.}
\label{linecut_highT}
\end{figure*}

\begin{figure*}[ht!]
\centering
\includegraphics[width= 7.2 in]{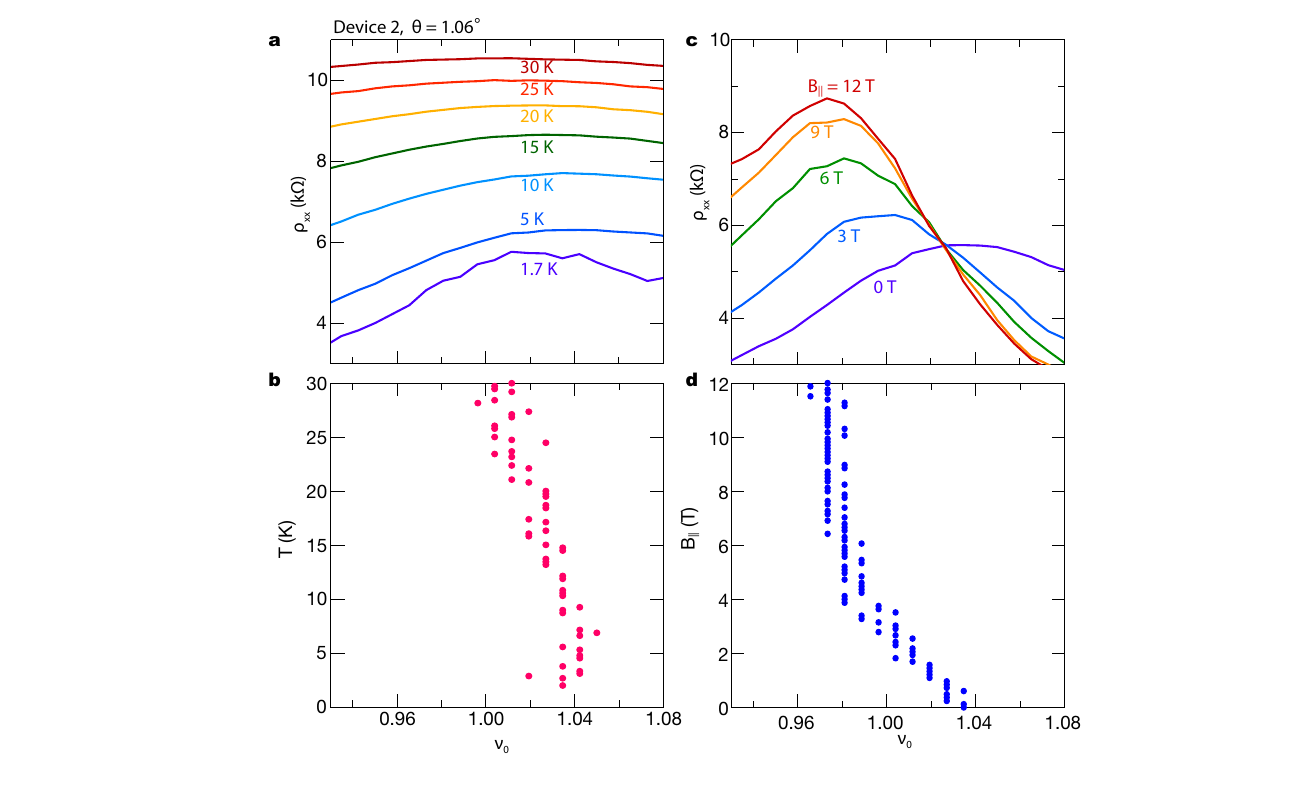}
 \caption{\textbf{Temperature and inplane magnetic field dependence of resistive peak around $\nu_0 = +1$ in Device 2.}
 \textbf{a}, $\rho_\mathrm{xx}$ as a function of nominal filling factor $\nu_0$ around $\nu_0 = +1$ between 1.7 and 30 K in Device 2. \textbf{b}, The  $\rho_\mathrm{xx}$ peak position as a function of $\nu_0$ and $T$.
 \textbf{c}, $\rho_\mathrm{xx}$ as a function of nominal filling factor $\nu_0$ around $\nu_0 = +1$ at $B_\parallel$ = 0, 3, 6, 9, 12 T. \textbf{b}, The  $\rho_\mathrm{xx}$ peak position as a function of $\nu_0$ and $B_\parallel$.}
\label{fig:XX40_peak}
\end{figure*}

\begin{figure*}[ht!]
\centering
\includegraphics[width= 7.6 in]{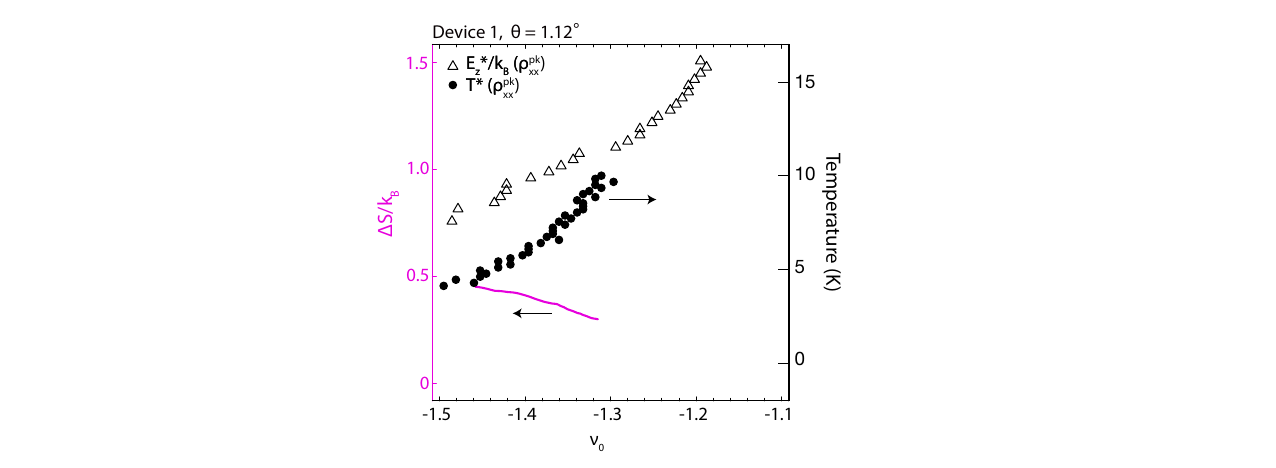}
 \caption{\textbf{Entropy change per superlattice unit cell $\Delta S/k_B$ from the transport data} White triangles and black circles are phase boundaries for Zeeman-tuned transition ($E^*_Z/k_B$) and temperature tuned transition ($T^*$), respectively, determined by the $\rho_\mathrm{xx}$ peak near $\nu = -1$.
 Pink curve is $\Delta s/k_B$ as a function of $\nu_0$ determined by Eq. (7).}
\label{S_transport}
\end{figure*}

\begin{figure}[ht!]
\includegraphics[width= 98.7mm]{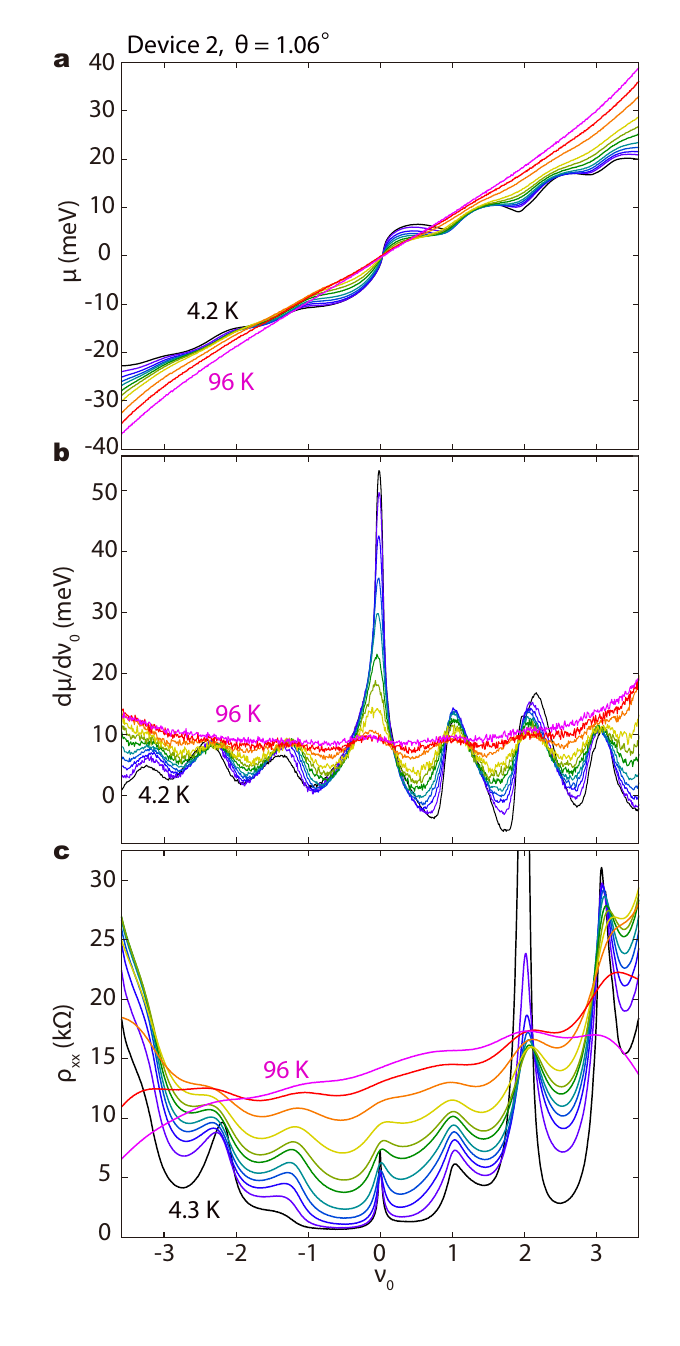}
\caption{\textbf{Temperature dependent chemical potential and resistance in Device 2.}
\textbf{a}, Chemical potential $\mu$ as a function of $\nu_0$ at 4.2, 8.0, 12, 16, 20, 26, 32, 40, 59, 76 and 96 K. $d\mu/d\nu_0$ in Fig.\ref{fig:4} of the main text is calculated by the derivative of these data.
\textbf{b}, Inverse electronic compressibility $d\mu/d\nu$ as a function of $\nu_0$ at 4.2, 8, 12, 16, 20, 26, 32, 40, 59, 76 and 96 K.
\textbf{c}, $\rho_\mathrm{xx}$ as a function of $\nu_0$ at 4.3, 8.0, 12, 16, 20, 26, 31, 40, 59, 76 and 96 K. 
}.
\label{Device2_highT}
\end{figure}

\begin{figure*}[ht!]
\centering
\includegraphics[width= 8.0 in]{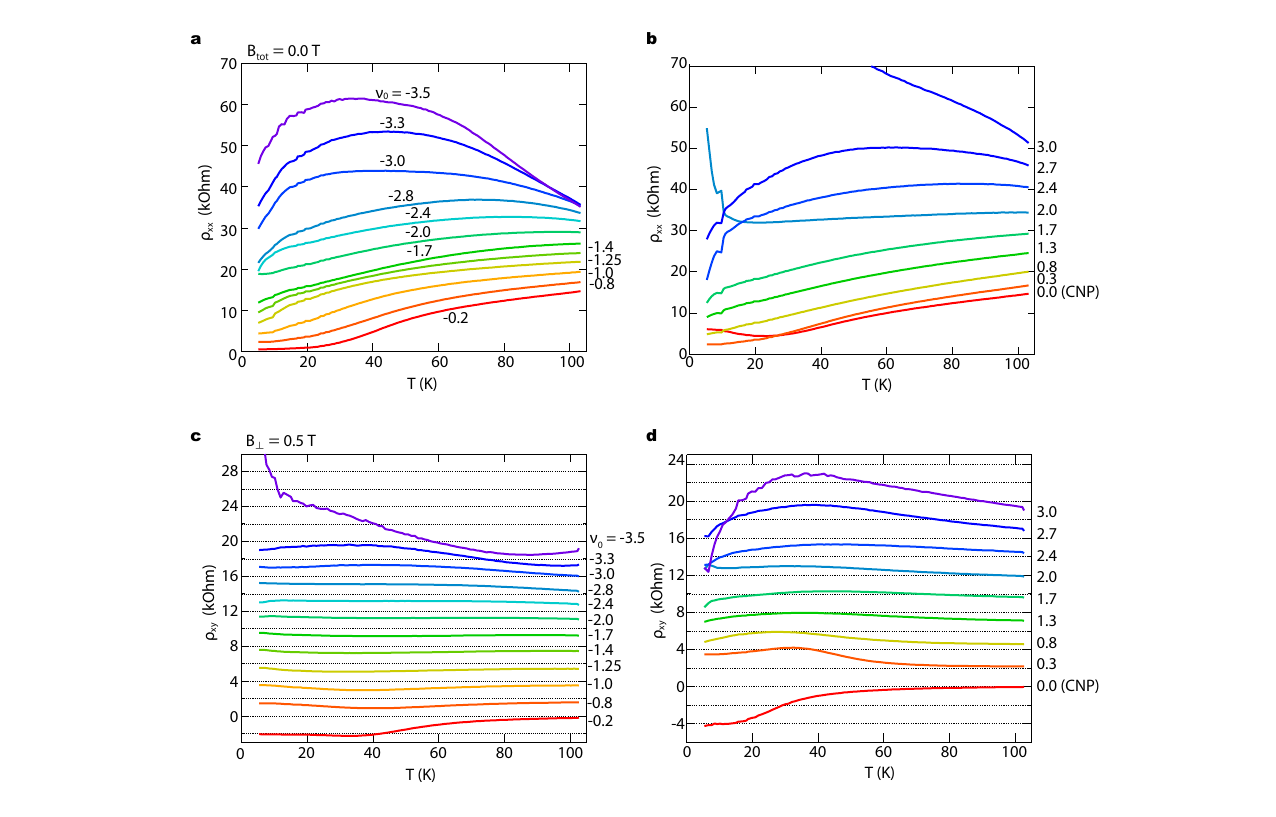}
 \caption{\textbf{Temperature dependence of $\rho_\mathrm{xx}$ and $\rho_\mathrm{xy}$ in Device 1.}
\textbf{a}, $\rho_\mathrm{xx}$ as a function of temperature at various $\nu_0$ for $\nu_0 <0$ and \textbf{b}, $\nu_0>0$. \textbf{c}, $\rho_\mathrm{xy}$ as a function of temperature at various $\nu_0<0$ and \textbf{d} $\nu_0 > 0$.  
All curves are offset by 2 k$\Omega$ for clarity.}
\label{RT}
\end{figure*}


\clearpage
\newpage 
\newpage 

\end{document}